%% file: main.tex
\documentclass[journal]{IEEEtran}
\usepackage[utf8]{inputenc}

\usepackage{xcolor}
\usepackage{color, soul}
\usepackage{enumerate}
\usepackage[shortlabels]{enumitem}
\usepackage{url}

\usepackage{hyperref}
\usepackage{cite}
\usepackage{bm, bbm} 
\usepackage{mathtools}
\usepackage{amsmath, amssymb, amsthm}
\usepackage{nicefrac}
\usepackage{empheq}

\usepackage{booktabs}
\usepackage{graphicx}
\usepackage{multirow}
\usepackage{units}
\usepackage{enumitem}
\usepackage{placeins}
\usepackage{threeparttable}
\usepackage{todonotes}

\ifCLASSOPTIONcompsoc
\usepackage[caption=false, font=normalsize, labelfont=sf, textfont=sf]{subfig}
\else
\usepackage[caption=false, font=footnotesize]{subfig}
\fi
\usepackage{hyperref}
\usepackage{svg}

\input{math_and_tools.tex}

\input{set_margins.tex}

\begin{document}

\bstctlcite{IEEE:BSTcontrol} 

\title{Weather-Driven Flexibility Reserve Procurement: \ \ A NYISO Offshore Wind Power Case Study}

\author{Zhirui~Liang,~
        Robert~Mieth,~
        Yury~Dvorkin,~
        and Miguel A. Ortega-Vazquez \vspace{-4mm}



}


\maketitle

\begin{abstract}
The growing penetration of variable renewable energy sources (VRES) requires additional flexibility reserve to ensure reliable power system operations. 
Current industry practice typically assumes a certain fraction of the VRES power production forecast as flexibility reserve, thus ignoring other relevant information, such as weather conditions. 
To address this, probability- and risk-based reserve sizing models have been proposed, which use probabilistic VRES power forecasts that mostly rely on historical forecast and actual VRES power data for model training.
Hence, these approaches are not suitable for planned or newly installed wind farms, where no or insufficient historical data is available.
This paper addresses this caveat. First, we propose a weather-driven probabilistic forecasting method for wind power installations using publicly available weather data. 
Second, we apply the resulting probabilistic forecasts to a novel risk-based flexibility reserve sizing model that is compatible with the current reserve procuring pipeline used by US ISOs. 
Finally, we compare the risk-based reserve requirements to industry practice, state-of-the-art reserve procurement methods, and a weather-ignorant benchmark with respect to system cost and security.
Our results are obtained from real-world data on a 1819-bus NYISO system model with both on- and projected off-shore wind power installations, which highlight the usefulness of weather information wind power forecasting and demonstrate efficiency gains from risk-aware reserve procurement.
\end{abstract}

\IEEEpeerreviewmaketitle

\section{Introduction}
\label{sec:introduction}

In current practice, flexibility reserve to compensate variability and uncertainty of variable renewable energy sources (VRES) is usually dimensioned and allocated in an \textit{ad hoc} manner under stationary assumptions, which may recognize a probabilistic nature of the VRES
\cite{dvorkin_2014,ela2011operating} but largely ignore \emph{current} ambient system conditions (e.g., weather) \cite{khatami2019flexibility}. Coupled with lacking weatherization of generation resources, overlooking weather-driven impacts on the size and allocation of flexibility reserve may lead to high-impact power outages (e.g., 2021 Texas power crisis, \cite{BUSBY2021102106}). Furthermore, the roll-out of large scale wind power plants, e.g., \unit[9]{GW} off-shore capacity in the NYISO service territory \cite{offshore_wind}, has increased the demand for flexibility reserve, which may in turn become scarce due to the recent or planned phase-out of fossil-fired units, e.g., \cite{nyiso_retirement}. To address these shortcomings, this paper internalizes weather-data statistics into the computation of reserve requirements and enables risk-aware flexibility reserve sizing and allocation within a day-ahead unit commitment decision process.

Existing approaches to quantify flexibility reserve requirements can be classified into two groups: (i) implicit approaches using stochastic optimization models and (ii) explicit approaches using deterministic optimization models \cite{ortega2020risk}. 
Implicit reserve sizing approaches internalize VRES uncertainty into a scheduling model. Some approaches rely on scenario-based stochastic programming, which solve a unit commitment problem over a set of representative scenarios and derive the optimal reserve requirement and allocation based on the scenarios and their probabilities \cite{zhang2013modeling,bruninx2016endogenous, 5739566}. 
However, the accuracy of the implicit approaches depends on the number and quality of the chosen scenarios, \cite{dupavcova2003scenario}, and requires extensive computational resources \cite{6872597}. 
To avoid scenario-based computations, \textit{robust optimization} approximates the scenarios by a predefined uncertainty set \cite{ning2019data} and derives flexibility reserve requirements to accommodate the worst-case VRES forecast error within a given set, which may increase the operating cost. 
To limit the cost increase, \textit{chance-constrained optimization} relaxes the robust approach by discarding some low-probability VRES outcomes \cite{bienstock_2004}, which may not guarantee the reserve adequacy for low-probability extreme scenarios. However, implementing the implicit reserve  approaches in practice is obstructed by their incompatibility with current market structures \cite{Miguel2021program}, which revolves around deterministic optimization.

Explicit reserve approaches, on the other hand, determine reserve requirements exogenously, relative to the scheduling optimization, and then enforce these requirements in the scheduling optimization. Most of the current reserve procurement methods are explicit and are \textit{extent-based}, i.e., the flexibility reserve should cover a certain percentage of forecast VRES power injection (or net load) \cite{Miguel2018reserve}, thus ignoring that the output and the forecast errors of VRES are not necessarily proportional.  In contrast, \textit{probability-based} reserve  methods account for a probability distribution of the VRES output and require reserve to compensate VRES forecast errors within a given confidence interval \cite{bruninx2014statistical,parker2020probabilistic}. 
While the extent- and probability-based approaches can internalize the historical VRES statistics, they ignore real-time impacts of VRES forecast deviations on the actual system conditions. \textit{Risk-based} reserve overcomes this gap by assessing the risk impact of each scenario, e.g., the risk of an VRES output forecast error is equal to its probability times its cost impact \cite{ortega2020risk,zhang2014convex,zhang2018risk,ghorani2019risk}. However, computing risk-based reserve requirements is complicated by the inability to exactly estimate  future system states, which requires computationally expensive and iterative risk exploration for different reserve allocations \cite{mieth2022risk}. 

Given current market designs and US ISO practice, it is likely that explicit reserve sizing and allocation approaches will remain the state-of-the-technology for the foreseeable future. Therefore, this paper develops an explicit approach enabling a risk-aware reserve procurement informed by VRES power forecasting.
This approach, and similar probability-based reserve determination methods as in \cite{zhang2013modeling,bruninx2016endogenous, 5739566,dupavcova2003scenario,ning2019data,bienstock_2004,bruninx2014statistical,parker2020probabilistic}, requires knowing a probability distribution of VRES forecast errors, which may not be readily available for power system operations that still largely rely on point forecasting of VRES power. In theory, probabilistic forecasting methods are also widely studied, e.g., in \cite{tahmasebifar2020new,wu2020probabilistic,xie2018nonparametric,kim2018short}, but are still only narrowly used in practice \cite{hobbs2022can}. For example, \cite{costilla2022operating} converts probabilistic forecasts into discrete scenarios suitable for scenario-based reserve procurement. 
However, most probabilistic forecasting methods rely on machine learning models, e.g., neural networks \cite{tahmasebifar2020new,wu2020probabilistic}, which are difficult to interpret and reproduce. Finally, these models require both forecast and actual site-specific VRES power data for training, so the trained models are limited in scalability for different VRES locations.

While analyzing historical VRES power injections is suitable for already operational wind and solar plants, such analyses are not possible for planned or newly installed systems. In this case, VRES power injections and their stochastic properties must be estimated from historical weather data. 
A straightforward way to transfer probabilistic forecasts of weather features (e.g., wind speed) to probabilistic forecasts of VRES power is by finding a functional relationship between weather and power output (e.g., power curve for wind turbines) and apply it to the former distribution. However, this functional relationship is complicated by various non-linearities \cite{7268773}. 
Moreover, modelling the compound influence of multiple weather features on VRES power (which is usually ignored in current practice) requires their joint distribution, which could be prohibitive since granular historical weather data is limited. 

To account for the effect of weather features on VRES (especially wind) power output, this paper adopts data stressing, a data science technique, to generates statistically credible wind power forecast errors that can be used to determine risk-based flexibility reserve requirements. First, instead of generating a continuous probability distribution of wind power, we generate statistically credible, stressed weather scenarios and map them into wind power scenarios accordingly. The weather scenarios are stressed by adding statistically consistent errors to the original weather forecasts using principal component analysis (PCA).
Second, instead of stressing all the weather features simultaneously, which may cause unnecessarily conservative forecast errors, one feature denoted as the \textit{key stressor} is stressed based on its historical distribution and other features are stressed according to their statistical properties in the original data, i.e., the correlation between different features.
After obtaining these scenarios, which collectively represent a range of potential wind power with corresponding probabilities, we can calculate the risk of each scenario and derive risk-based reserve requirements. 

The main contributions of this paper are: 
\begin{itemize}
    \item In contrast with \cite{zhang2013modeling,bruninx2016endogenous, 5739566,dupavcova2003scenario,ning2019data,bienstock_2004,bruninx2014statistical,parker2020probabilistic}, which directly model the uncertainty of wind power, this paper models weather uncertainty and maps it into uncertain wind power generation and quantifies the real-time scheduling cost savings of the proposed weather-driven method relative to the weather-ignorant benchmark.
    \item Instead of using machine learning models as in \cite{tahmasebifar2020new,wu2020probabilistic}, this paper designs a transparent, interpretable, and reproducible data stressing method to generate probabilistic wind power forecasts based on publicly available weather data. This also allows for applications at prospective wind farm sites where no historical wind output power data is available.
    \item We propose a risk-based reserve sizing and allocation procedure that avoids expensive probability computations as in \cite{ortega2020risk,zhang2014convex,zhang2018risk,ghorani2019risk} by leveraging the discrete nature of our scenarios. This procedure follows the current reserve determination pipeline used by US ISOs, which eases adoption in practice. This reserve procurement procedure is also compatible with other probabilistic forecasting methods. 
    \item We demonstrate the effectiveness of the proposed weather-driven method and risk-based reserve procurement on a 1819-bus New York Independent System Operator (NYISO) transmission network model and real weather data. We show the proposed  approach is compatible with existing system-wide, zonal, and nodal reserve policies, as well as contingency reserve requirements.
    As a result, we consider the numerical conclusions obtained in this paper to be also of interest to system operators and policymakers working towards future sustainable power systems.
\end{itemize}

\section{Weather-Driven Wind Uncertainty Model}
\label{sec:wind}
Efficient procurement of flexibility reserve, e.g., in day-ahead planning and market-clearing procedures, requires quantifying potential differences between wind power forecasts and real-time injections, i.e., the forecast errors of wind power. 
This section proposed a weather-driven method of generating probabilistic forecast of wind power based on its point forecast. The weather features and their potential forecast errors are correlated \cite{pu2019numerical} and their compound effect on the wind power output and its forecast error must be considered. Realizing that wind speed is the key driver and thus the key stressor for available wind power, Section~\ref{subsec:stress_wind_speed} describes a method to generate wind speed forecast errors from historical distributions.
Then, Section~\ref{subsec:stress_other_weather}, describes a method to generate forecast errors of other weather features accordingly, while maintaining the correlation between different weather features. 
Finally, Section~\ref{subsec:stress_wind_power} translates these stressed multi-feature weather scenarios into stressed wind power scenarios, which informs the reserve sizing and allocation procedure in Section~\ref{sec:reserve}.

The application of the proposed method is illustrated using  data from the planned  ``Empire2'' Offshore Wind Power Plant in New York, USA \cite{Empire2_Wind}, with the total capacity of \unit[1260]{MW}, the individual turbine capacity of \unit[15]{MW}, and the turbine hub height of \unit[100]{m} (as in the turbines from the supplier selected for this project \cite{vestas_empire}).
Since this wind farm is still under construction and no output power data is available, we use day-ahead forecast wind speed (1-hour resolution) and real-time wind speed (5-min resolution) at the height of \unit[100]{m} from the Wind Integration National Dataset (WIND) Toolkit of NREL \cite{Wind_toolkit}. The WIND Toolkit also provides real-time values (5-min resolution) for 32 additional weather features, including the air pressure, humidity, temperature, and wind speed/direction at different altitudes.

\subsection{Weather Data for Wind Power Calculations}
\label{subsec:weather_data}
\begin{figure}[!t]
    \centering
    \includegraphics[width=1\linewidth]{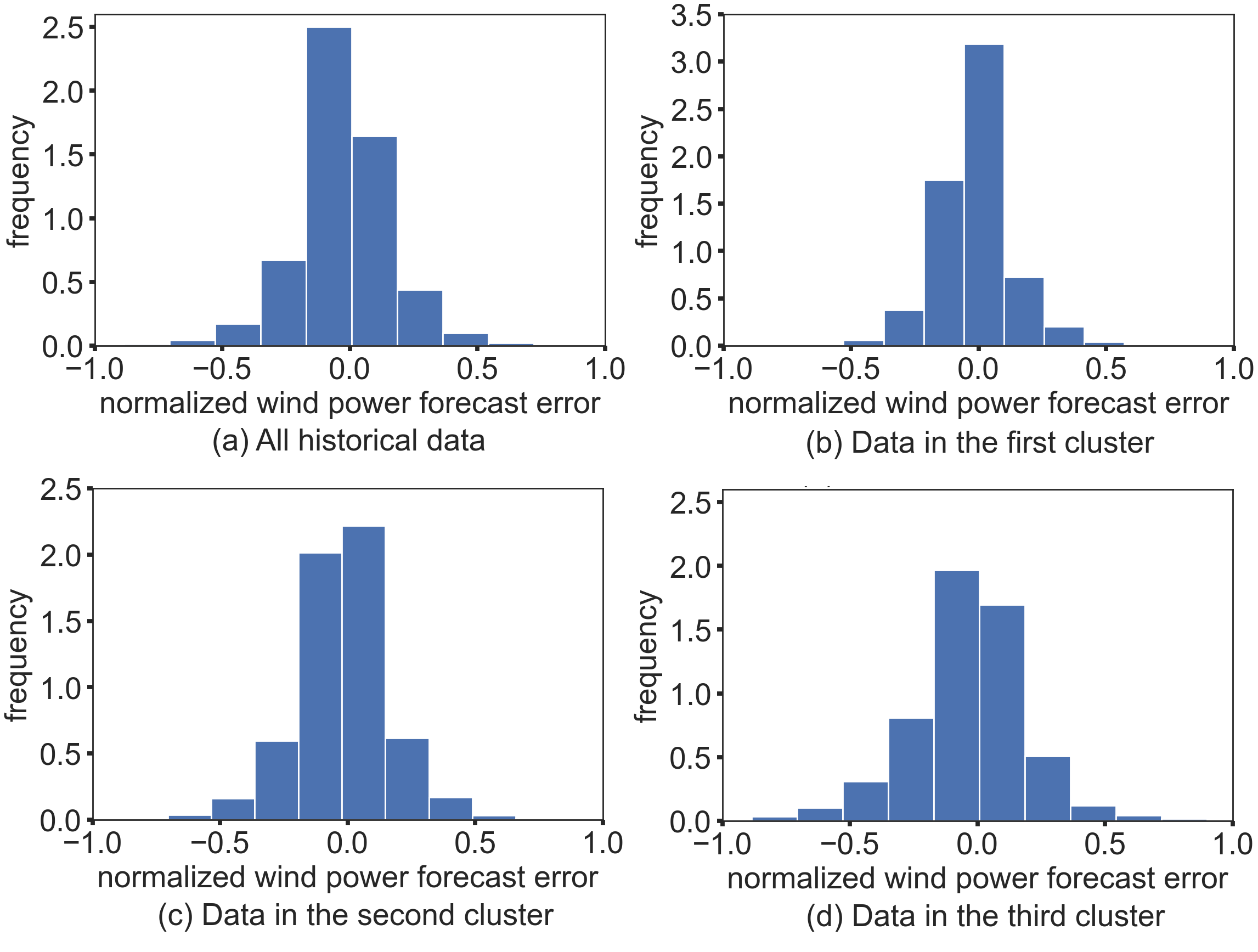}
    \caption{Historical distribution of wind power forecast error based on (a) all historical data (resource: NREL WIND Toolkit \cite{Wind_toolkit}); (b), (c), (d) historical data in the first, second and third clusters. Each cluster of data corresponds to a unique weather pattern identified with the k-means algorithms.}
    \label{fig:cluster}
\end{figure}

Weather conditions has impact on the distribution of wind power forecast errors. Fig.~\ref{fig:cluster} compares the historical distribution of wind power forecast errors of the ``Empire2'' wind farm based on all historical data and historical data in three clusters. The clusters are obtained by applying K-means clustering on the time series of five weather features, including wind speed, wind direction, air temperature, humidity, and air pressure. Each cluster of data corresponds to a unique weather pattern. According to Fig.~\ref{fig:cluster}, the error distributions are different for different weather patterns, which shows the necessity to incorporate weather information into wind power forecasting.

The power output of a single wind turbine ($P_{\rm wind}$) can be approximated as:
\begin{align}
    P_{\rm wind}=\frac{1}{2}\rho A_{\rm rotor} V_{\rm wind}^3 C_{p}, \label{power_calculation}
\end{align}
where $\rho$ is the air density, $A_{\rm rotor}$ is the rotor swept area, $V_{\rm wind}$ is the wind speed, and $C_{p}$ is the power coefficient, which denotes a recoverable fraction of kinetic wind power into electric power and ultimately depends on wind speed $V_{\rm wind}$ \cite{sohoni2016critical}.

\begin{figure}[!t]
    \centering
    \includegraphics[width=0.8\linewidth]{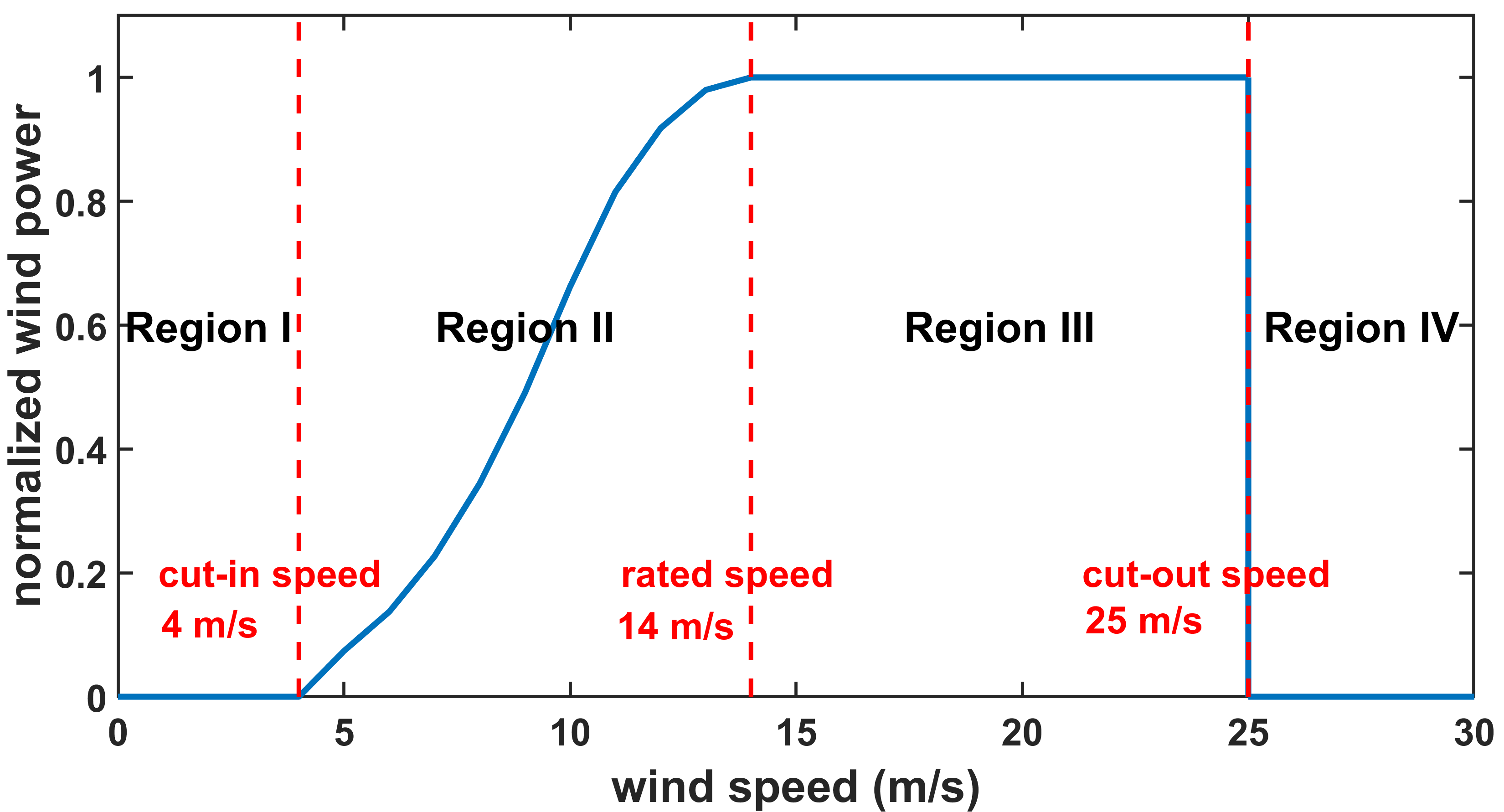}
    \caption{Normalized power curve of the NREL reference offshore wind turbine \cite{Power_curve}. Wind speeds are divided to four regions by the cut-in speed (when the blades start to rotate and generate power), rated speed (when the turbine starts to generate power at its maximum capacity) and cut-out speed (when the turbine must be shut down to avoid damage to the equipment).}
    \label{fig:power_curve}
\end{figure}

Assuming the values of $\rho$ and $A_{\rm rotor}$ are constant and the curve of $C_{p}$ with respect to $V_{\rm wind}$ is given, the relationship between $V_{\rm wind}$ and $P_{\rm wind}$ in \cref{power_calculation} can be simplified to the theoretical wind power curve shown in Fig.~\ref{fig:power_curve} \cite{Power_curve}. 
However, the actual relationship between wind speed and wind power does not always follow this theoretical curve, because it ignores (i) the impacts of other weather features, e.g., humidity and air density \cite{zhang2014review, astolfi2021perspectives}, (ii) the wake effects from neighboring wind turbines \cite{stevens2016effects}, and (iii) the combined effect of wind speed at different altitudes \cite{saint2020parametric}. 
To better reflect the relationship between wind power and other weather conditions, we modify \cref{power_calculation} as follows:

\noindent
\underline{\textit{Step 1}}: 
Instead of using a constant air density $\rho$ (usually set to \unit[1.225]{kg/m$^3$} \cite{saint2020parametric}), we calculate air density at the hub height ($\rho_{\rm hub}$) using air pressure ($P_{\rm hub}$), humidity ($H_{\rm hub}$), and air temperature ($T_{\rm hub}$) at the hub height based on Eqs.~(2)--(3) from \cite{jung2019role}. 

\noindent
\underline{\textit{Step 2}}: 
We assume that the loss due to internal and external wake effects is $\delta=15\%$. Based on $\delta$ and $P_{\rm hub}$, we can obtain the modified normalized power curve as:
\begin{align}
    f_{\rm mod}(V_{\rm wind})=(1-\delta)\frac{1}{2}\rho_{\rm hub} A_{\rm rotor} V_{\rm wind}^3 C_{p}. \label{power_curve_modified}
\end{align}

\begin{figure}[!t]
    \centering
    \includegraphics[width=0.8\linewidth]{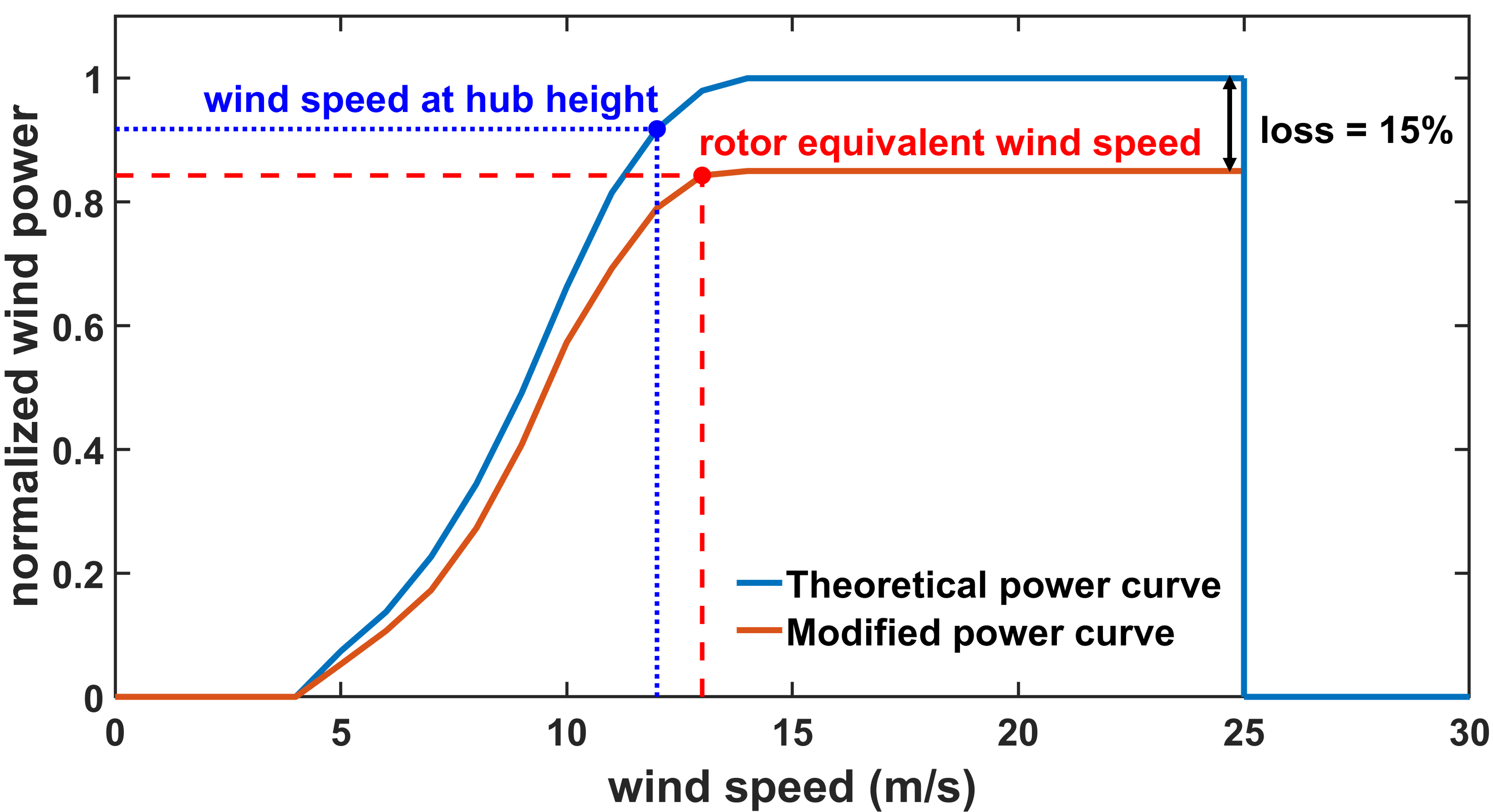}
    \caption{Wind power calculation based the modified power curve and the rotor equivalent wind speed.}
    \label{fig:power_curve_modified}
\end{figure}

\noindent
\underline{\textit{Step 3}}:
Instead of using wind speed at the hub height (i.e., the blue dot in Fig.~\ref{fig:power_curve_modified}), we use the rotor equivalent wind speed $V_{\rm equ}$ (i.e., the red dot in Fig.~\ref{fig:power_curve_modified}) calculated from a collection of wind speeds and directions at different altitudes using Eqs.~(7)--(9) from \cite{saint2020parametric}. In this paper, we use 9 sets of wind speed and wind direction from \unit[40]{m} to \unit[200]{m}. The normalized output power of the wind farm can then be calculated by substituting $V_{\rm equ}$ into the modified power curve \cref{power_curve_modified} as:
\begin{align}
    P_{\rm norm}=f_{\rm mod}(V_{\rm equ}), \label{power_calculation_modified}
\end{align}
and the actual output power of the wind farm is $P_{\rm norm}S_{\rm farm}$, where $S_{\rm farm}$ is the installed capacity of the wind farm.

In total, this paper uses 21 weather features, itemized in Table~\ref{tab:21_features}, to calculate the wind power output. More/less features could be used in practice to meet actual data availability.
For example, turbulence intensity could be added to the model based on the method introduced in \cite{saint2020parametric} without modifying the proposed method in subsequent sections.

\begin{table}[!t]
  \centering
  \caption{Weather Features Used in Wind Power Calculation}
    \begin{tabular}{cl}
    \toprule
    No.	& Weather Feature	 \\
    \midrule
    1	& air pressure at \unit[100]{m} (Pa)	\\
    2	& relative humidity at \unit[2]{m} (\%)		 \\
    3	& air temperature at \unit[100]{m} ($^{\circ}$C)	\\
    4--12 & wind direction at \unit[40, 60, 80, ..., 200]{m} ($^{\circ}$) \\
    13--21 & wind speed at \unit[40, 60, 80, ..., 200]{m} (m/s)	\\
    \bottomrule
    \end{tabular}%
  \label{tab:21_features}
\end{table}

\subsection{Stressed Scenarios for Wind Speed}
\label{subsec:stress_wind_speed}

We analyze wind speed for four regions (or intervals) of the non-linear wind turbine power curve shown in Fig.~\ref{fig:power_curve}.
Within Regions~I, III, and IV, the power output of the turbine is independent of the exact wind speed as it is either zero (when wind speed is below the cut-in or above cut-off speed) or at the rated level.
Only wind speed values in Region~II require wind power calculations.
Therefore, forecast wind power in day-ahead and actual wind power in real-time differ only if (i) forecast and actual wind speeds are both in Region~II, (ii) forecast and actual wind speeds are in different regions. 
In case of (i), a small forecast error in wind speed can lead to a large error in wind power due to the cubic relationship between wind speed and wind power.
In case of (ii), the wind power forecast error could be even greater, e.g., if the actual wind speed exceeds the cut-out speed and moves from Region~III to Region~IV.
To capture this behavior, we design a two-step approach to model the distribution of wind speed forecast errors by first modeling transition probabilities between Regions~I--IV and, second, by mining a set of conditional distributions on wind speed forecast errors for each region.

\subsubsection{Transition Matrix}
\begin{figure}[!t]
    \centering
    \includegraphics[width=0.9\linewidth]{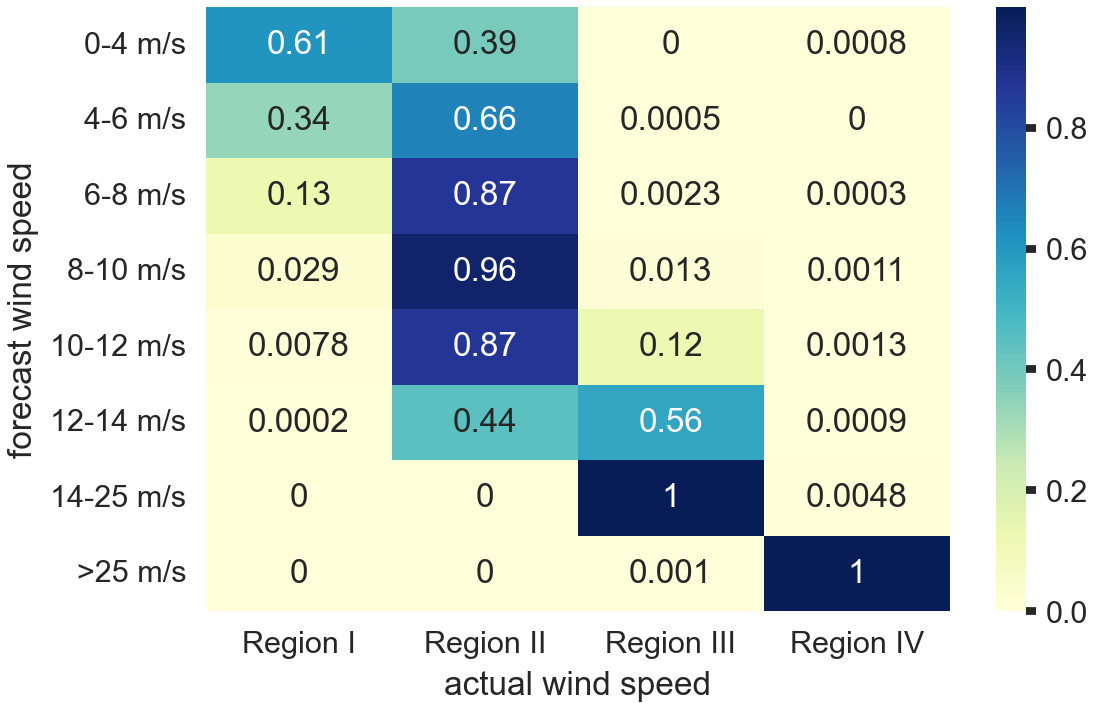}
    \caption{Transition matrix between forecast wind speed itemized in intervals and actual wind speed itemized in regions of the power curve (see Fig.~\ref{fig:power_curve}).}
    \label{fig:transition_matrix}
\end{figure}

For each time step in the available data set, we create a pair $(V^{F}, V^{A})$ of forecast wind speed $V^{F}$ and actual wind speed $V^{A}$.
While each $V^{A}$ is assigned its respective region of the power curve, we assign $V^{F}$ to higher resolution intervals to increase modeling fidelity. The y-axis of Fig.~\ref{fig:transition_matrix} itemizes these intervals.
Now, each of these tuples can be assigned a transition between wind speed interval $m$ and region $n$.
Using the historical data, we compute transition probability $P_{mn} = P({V^{A}} \in {\rm{Region}}\ n | {V^{F}} \in {\rm{Interval}}\ m),\ m \in \set{I},\ n \in \{\rm{I,II,III,IV}\}$ by dividing the number of transitions between each $mn$-pair by the total number of transitions. 
These transition probabilities are organized in a transition matrix, see Fig.~\ref{fig:transition_matrix}, where the value of each cell is the transition probability, and the row-wise sum of transition probabilities is one. 
For example, if the forecast wind speed is in the interval \unit[0-4]{m/s}, then the likelihood that the actual wind speed is in Regions~I and II is about 61\% and 39\%, respectively, and the likelihood of the actual wind speed to appear in Regions~III and IV is negligible. 

\subsubsection{Conditional Distributions}
\begin{figure}[!t]
    \centering
    \includegraphics[width=1\linewidth]{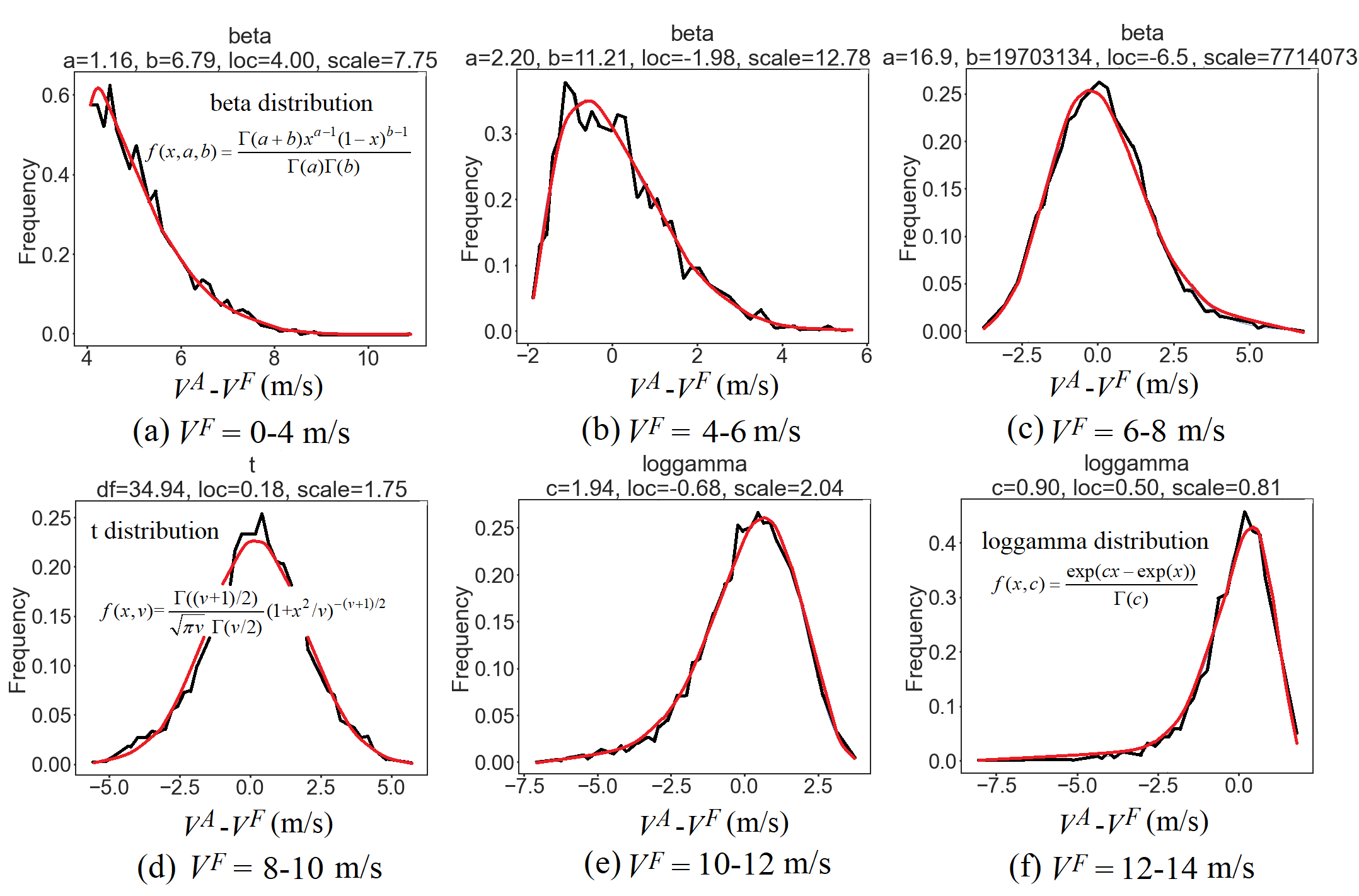}
    \caption{Distribution of wind speed forecast errors in different intervals. The black lines are the empirical distributions, and the red curves are the best-fit parametric distributions. The type and parameters of each distribution are shown above each sub-figure. Equations in (a), (d), (f) are the corresponding probability density functions.}
    \label{fig:error_distribution}
\end{figure}

Whenever the probability of the actual wind speed  in Region~II is non-zero, we model the specific distribution of the forecast errors. 
First, we compute the historical wind speed forecast error as $V^{A} - V^{F}$ and then fit suitable \textit{forecast error} distributions conditioned by the forecast wind speed. 
The resulting conditional error distributions are shown in Fig.~\ref{fig:error_distribution} for relevant wind speed intervals.
We use the Python package \textit{distfit} \cite{Taskesen_distfit} to find the best-fit functions as reported in Fig.~\ref{fig:error_distribution}.

\subsubsection{Error Sampling}
Using the transition matrix and the conditional distributions, we sample forecast errors for a given wind speed forecast. 
For example, assume the forecast wind speed for a time interval is \unit[9]{m/s} and we want to sample 1000 wind speed forecast errors. 
First, as per the fourth row in the transition matrix in Fig.~\ref{fig:transition_matrix}, the 1000 samples should be distributed between Regions~I, II, III, and IV as 29, 957, 13, and 1, respectively. 
Second, the 957 points in Region~II should follow the distribution in Fig.~\ref{fig:error_distribution}(d), while the points in the other three regions can be set to constant values, which we choose as \unit[1]{m/s}, \unit[20]{m/s}, and \unit[30]{m/s} without loss of generality.

\begin{figure}[!t]
    \centering
    \includegraphics[width=1\linewidth]{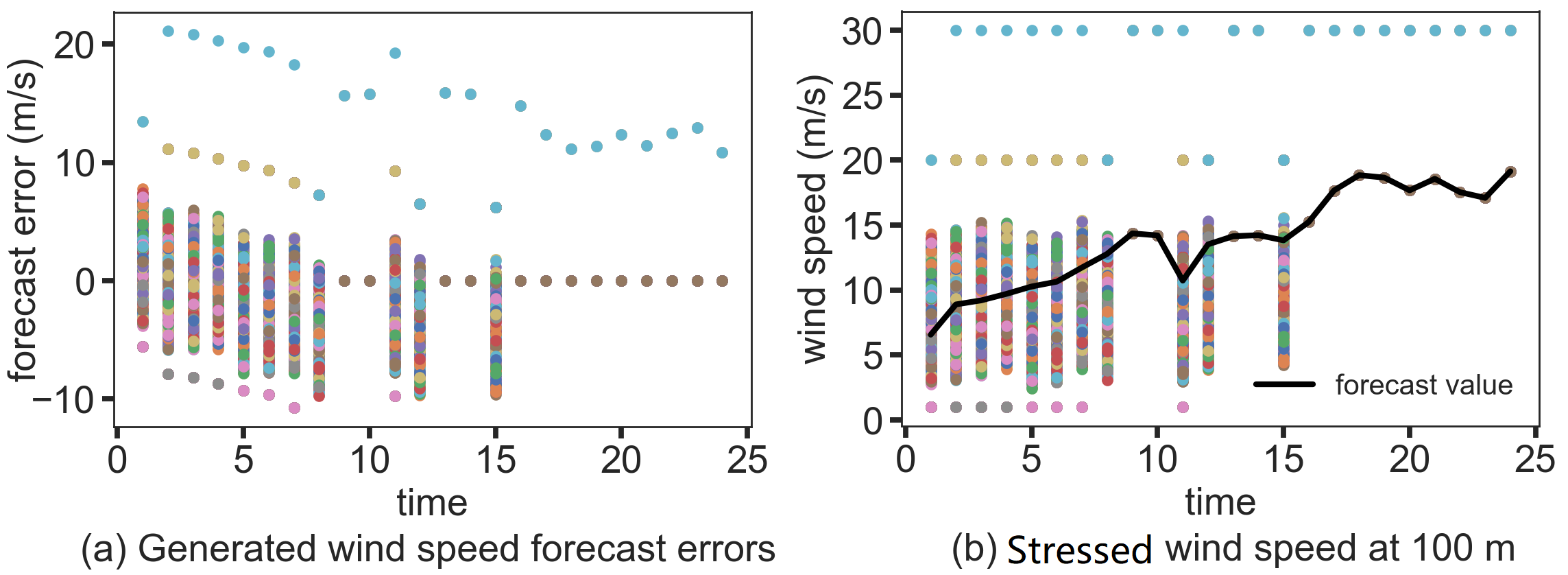}
    \caption{Stressed scenarios for wind speed at \unit[100]{m}.}
    \label{fig:speed_errors}
\end{figure}

For illustration purposes, we perform a numerical experiment using data from Jan. 30, 2013. 
The generated wind speed forecast errors are shown in Fig.~\ref{fig:speed_errors}(a). 
By adding these errors to the forecast 24h wind speed profile, we obtain statistically credible real-time wind speed scenarios, which we call \textit{stressed scenarios for wind speed}, as shown in Fig.~\ref{fig:speed_errors}(b). 
It can be seen that when the forecast wind speed is in Region~II (e.g., before hour 8), the distribution of the stressed wind speeds is spread across all four regions. Further, when the forecast wind speed is in Region III (e.g., after hour 16), the stressed wind speeds are concentrated in two values, i.e., the forecast wind speed at the respective hour and \unit[30]{m/s}, because the actual wind is expected to fall into Regions~III or IV almost surely.

\subsubsection{Discussion}
The procedure above assumes that the wind speed forecast errors are only related to the forecast wind speed, while the effect of other factors, e.g., inter-temporal correlations, are ignored. 
This effect can be taken into account by using rolling wind speed forecasting or by quantifying statistical parameters of scenarios spanning across  multiple resolutions, \cite{7268773,ortega2022generation}.

Relative to the method in \cite{zhang2013modeling}, which models a conditional distribution of wind power forecast errors for possible power outputs, no advanced calculation methods are used and fewer cases need to be considered. This simplification is achieved by leveraging the properties of the wind  power curve, i.e., realizing that changes in wind speed will affect wind power only when wind speed values fall in Region~II or transition between regions.

\subsection{Stressed Scenarios for Other Weather Features.}
\label{subsec:stress_other_weather}

\begin{figure}[!t]
    \centering
    \includegraphics[width=0.75\linewidth]{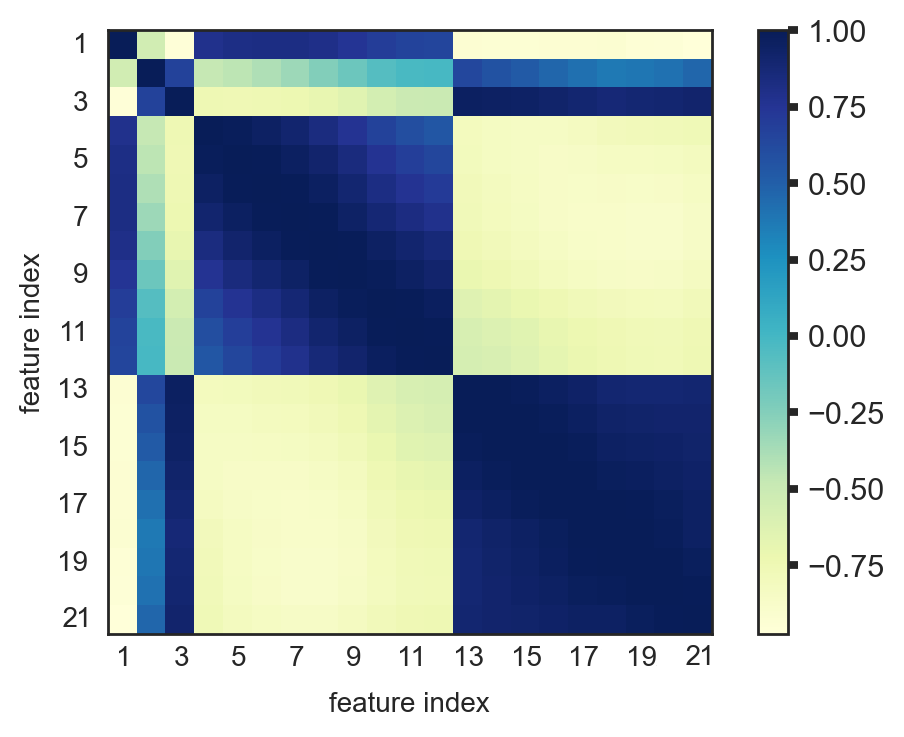}
    \caption{Correlation matrix for the 21 weather features in Table~\ref{tab:21_features}.}
    \label{fig:correlation}
\end{figure}

To keep the statistical properties of multi-feature weather data, weather features other than wind speed must be modified to match the stressed wind speed. 
Fig.~\ref{fig:correlation} shows the correlation matrix of 21 weather features for the data from Jan. 30, 2013. The wind speeds (or wind directions) at different altitudes are  positively correlated, while the wind speed and wind direction are negatively correlated. However, it is hard to stress all the weather forecast data while maintaining the correlation between different weather features due to the high dimension of this correlation matrix. 
Therefore, we use PCA to reduce the correlation matrix to a lower-dimensional linear relationship between the 21 weather features. 
With this relationship, we can generate the forecast errors of weather features other than the key stressor (in this case, the wind speed at \unit[100]{m}) based on the forecast errors of the key stressor. 

PCA is a statistical procedure widely used for dimensionality reduction, increasing data interpretability, and at the same time, minimizing information loss, by creating new uncorrelated variables (so-called principal components -- PCs). We refer the interested reader to \cite{jolliffe2016principal,daffertshofer2004pca} for more details on PCA. 
The first PC is the eigenvector of maximum variance of the original data, while the second PC represents the direction of second highest variance, and so on. 

PCA is applied to the data of weather features as follows:

\noindent
\underline{\textit{Step 1}}: Construct  standardized weather data matrix $X$ with dimension $h \times p$, where $h$ is the number of time steps (hours in a day, i.e., 24), and $p$ is the number of considered weather features (21 in this paper). 
We normalize wind direction using the sine function, which also removes the discontinuity between 359$^{\circ}$ and 0$^{\circ}$.

\noindent
\underline{\textit{Step 2}}: Calculate the covariance matrix of $X$ as $C = X^TX$.
Note that covariance matrix $C$ of  standardized data $X$ is the correlation matrix of the original data. 
PCA on the standardized data is also known as correlation matrix PCA \cite{jolliffe2016principal}. Correlation matrix PCs are invariant to linear changes in units of measurement and are therefore the appropriate choice for datasets where variables have different scales \cite{jolliffe2016principal}.

\noindent
\underline{\textit{Step 3}}: Compute the eigenvectors of $C$ and sort them in descending  order  of eigenvalues. The eigenvectors are referred as the PC loadings, which are the coefficients for the linear combination used to calculate individual PCs. 
The variance of the original data that each PC accounts for is given by its corresponding eigenvalue \cite{daffertshofer2004pca}.

\noindent
\underline{\textit{Step 4}}: Keep the first $k$ eigenvectors of $C$ ($k \le p$) and compute reconstruction $\tilde{X} = \sum\nolimits_{i = 1}^k {{\lambda_i}} {V_i}$, where $\lambda_i$ is the $i^{th}$ eigenvalue and $V_i$ is the $i^{th}$ eigenvector.

\noindent
\underline{\textit{Step 5}}: Restore $\tilde{X}$ to the scale of the original data by performing the inverse operation of the standardization in Step 1. After that, we obtain the coefficients of a linear relationship between different weather features.

\begin{figure}[!t]
    \centering
    \includegraphics[width=1\linewidth]{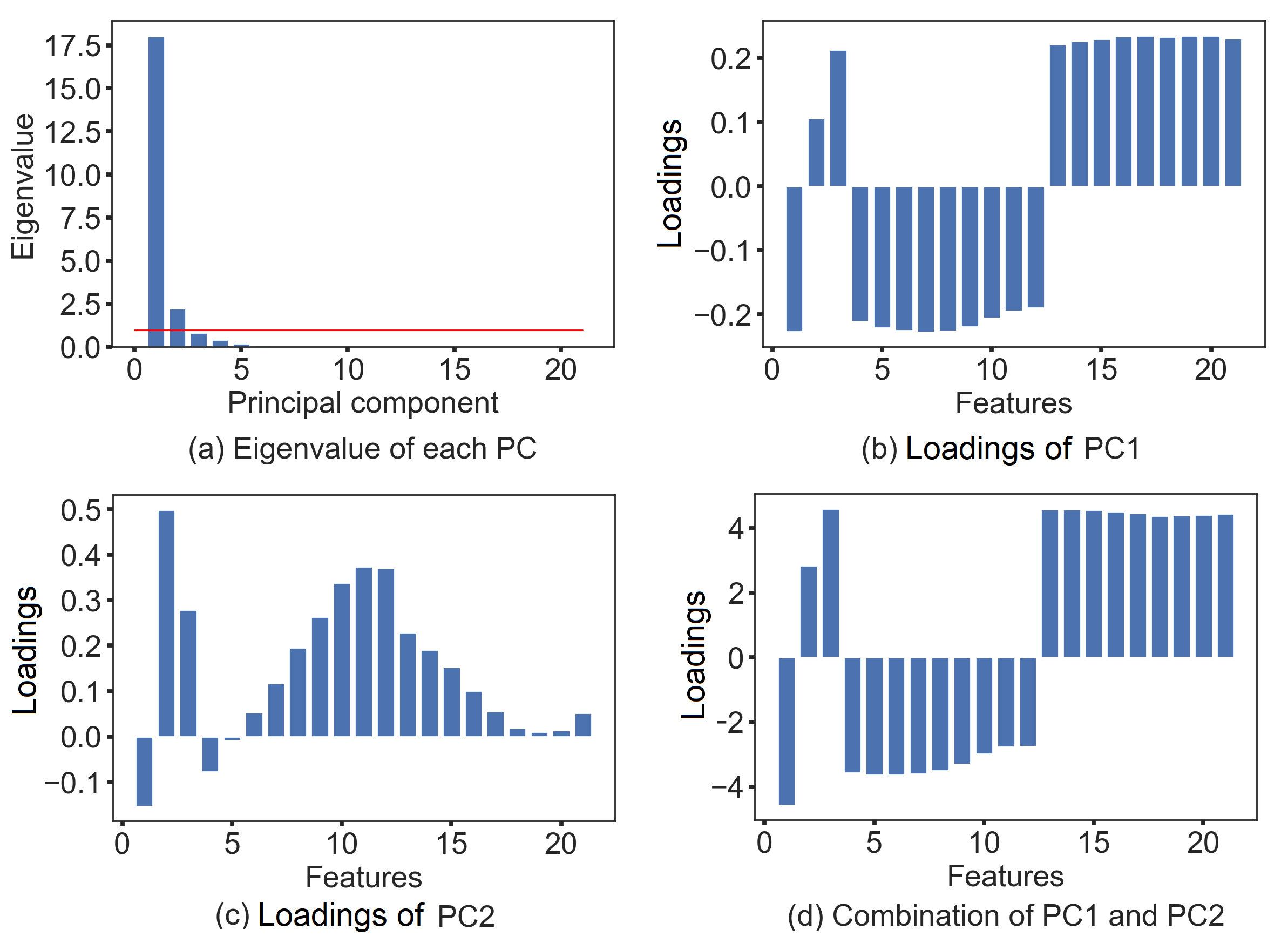}
    \caption{Eigenvalues and eigenvectors (loadings) of the correlation matrix. 
    (a) The eigenvalues of each PC; (b), (c) The PC loadings of the first and second PC; (d) The weighted sum of the first and second PC loading.}
    \label{fig:PCs}
\end{figure}

Again, for illustration, we perform this five-step procedure for the data on Jan 30, 2013.
Fig.~\ref{fig:PCs}(a) shows the resulting eigenvalues in a descending order.
We focus on the first two PCs whose corresponding eigenvalues are greater than one as per Kaiser's rule, a common, ad-hoc rule for selecting principal components \cite{kaiser1960application}.
The loadings of these PCs are shown in Fig.~\ref{fig:PCs}(b) and (c). 
We observe that in the first PC, the wind speed and the wind direction are negatively correlated, while in the second PC, they are positively correlated. The combination (weighted with the corresponding eigenvalues) of the first two PCs is shown in Fig.~\ref{fig:PCs}(d). 
In total, an increase in wind speed is correlated to a ``decrease'' in wind direction and air pressure, and an increase in humidity and temperature. 

\begin{figure}[!t]
    \centering
    \includegraphics[width=1\linewidth]{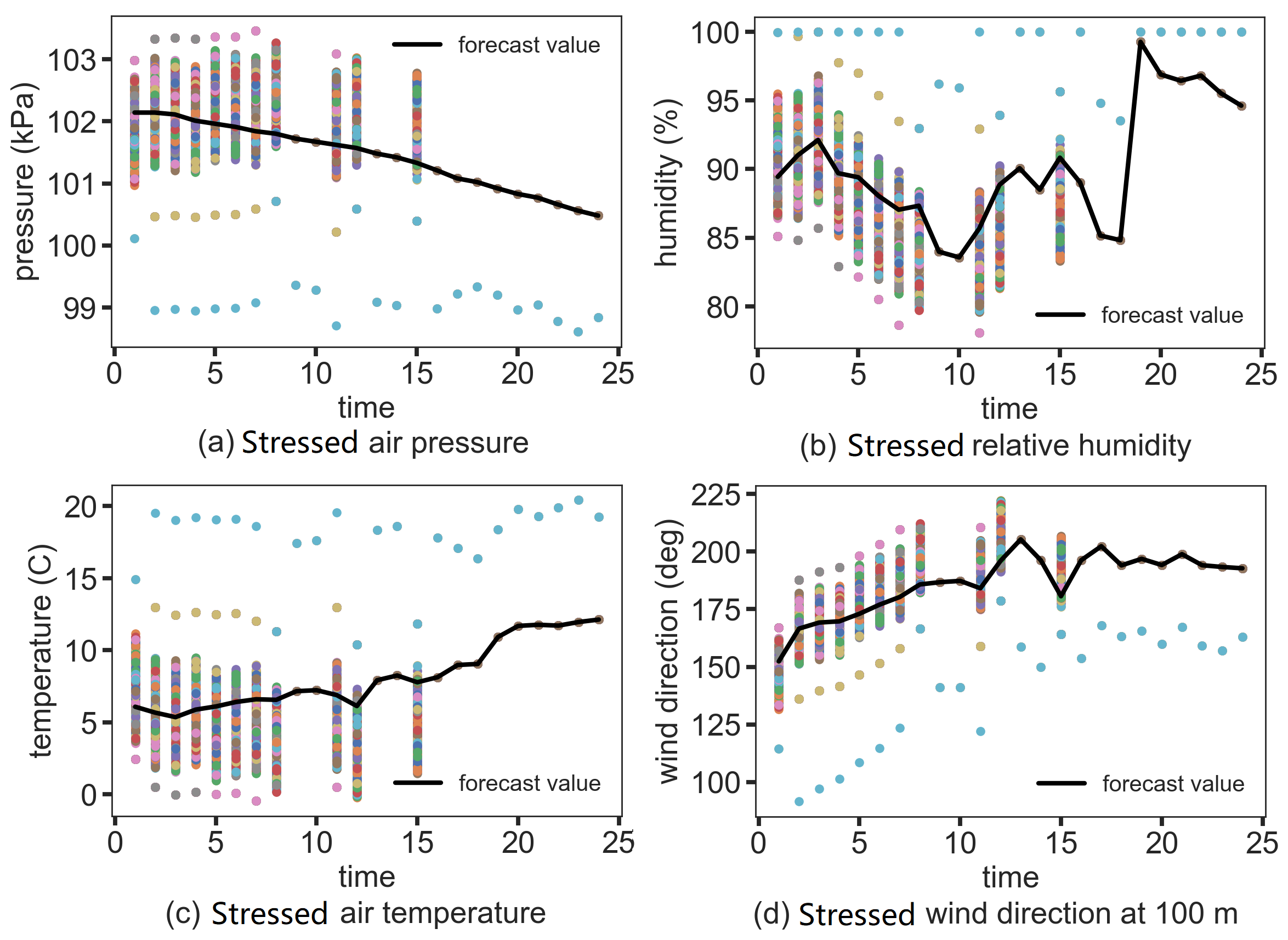}
    \caption{Stressed scenarios for four representative weather features.}
    \label{fig:stressed_weather}
\end{figure}

The coefficients obtained after Step 5 are shown in Table~\ref{tab:denormalized_PCA}, which represents the linear relationship between wind speed and other weather features. For example, when the wind speed at \unit[100]{m} increases by \unit[1]{m/s}, the air pressure is expected to decrease by \unit[15.9445]{pa} and the relative humidity is expected to increase by \unit[0.7817]{\%}. 
We use these results to generate 1000 statistically consistent scenarios for the remaining 20 weather features based on the stressed wind speeds in Fig.~\ref{fig:speed_errors}(b).
Fig.~\ref{fig:stressed_weather} shows the result for four representative weather features.

\begin{table}[!t]
  \centering
  \caption{Linear relationship between different weather features}
    \begin{tabular}{c c | c c| c c}
    \toprule
    No. & Coefficient & No. & Coefficient & No. & Coefficient\\ 
    \midrule
    1  & -15.9445 & 8  & -0.0463 & 15  & 0.9853 \\ 
    2  & 0.7817  & 9  & -0.0457  & 16  & 1.0000 \\ 
    3  & 0.6567  & 10  & -0.0455  & 17  & 1.0116 \\ 
    4  & -0.0521  & 11  & -0.0462  & 18  & 1.0373 \\ 
    5  & -0.0490  & 12  & -0.0507  & 19  & 1.0821 \\ 
    6  & -0.0474  & 13  & 1.0087  & 20  & 1.1771 \\
    7  & -0.0468  & 14  & 0.9976  & 21  & 1.2195 \\
    \bottomrule
    \end{tabular}%
  \label{tab:denormalized_PCA}
\end{table}

\subsection{Probabilistic Forecasting of Wind Power}
\label{subsec:stress_wind_power}

\begin{figure}[!t]
    \centering
    \includegraphics[width=1\linewidth]{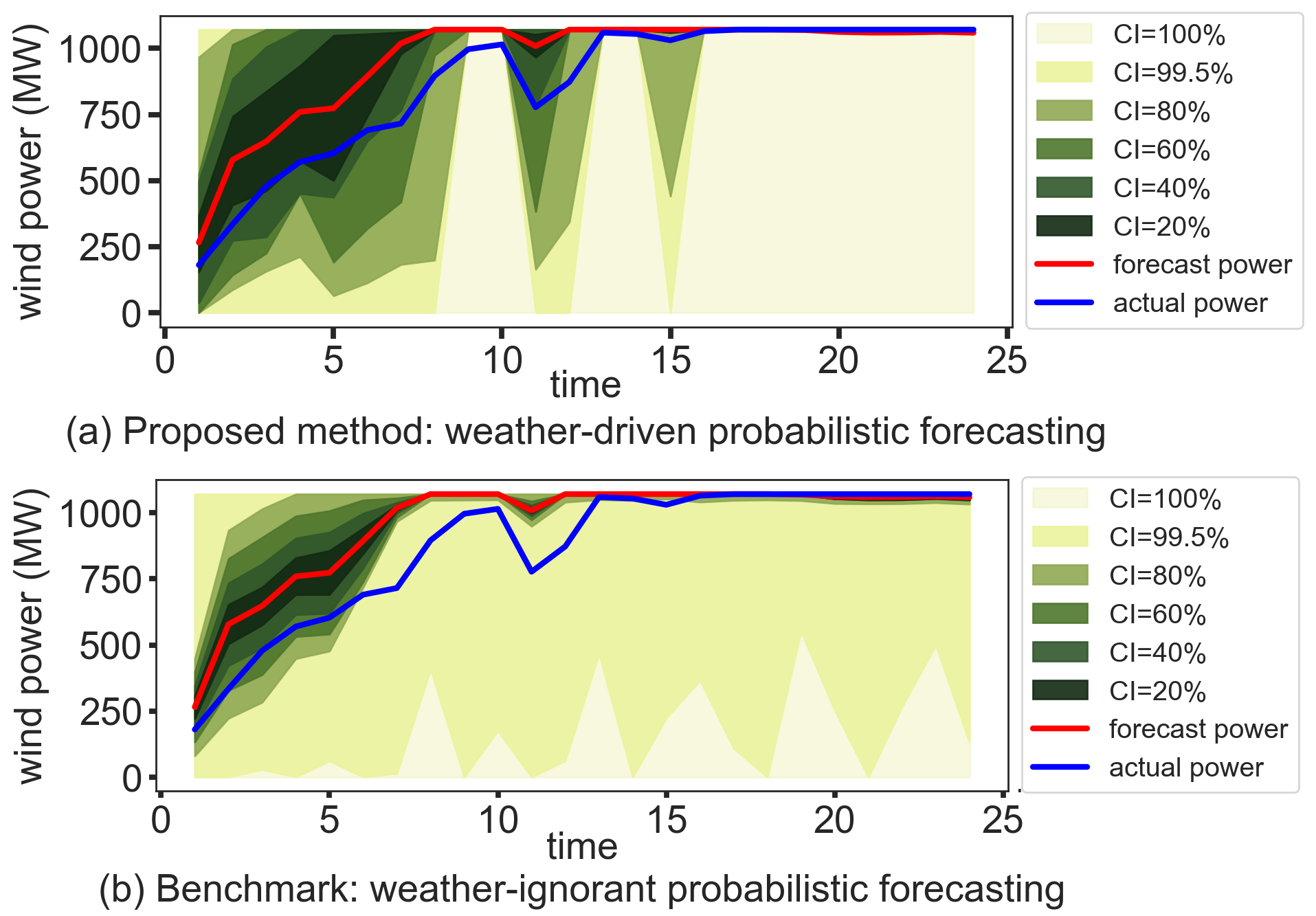}
    \caption{Stressed wind power with confidence intervals (CI). The label of ``CI=100\%'' means considering all the 1000 generated errors in Fig.~\ref{fig:speed_errors} and Fig.~\ref{fig:stressed_weather}, while ``CI=99.5\%'' means ignoring the 0.5\% largest errors in absolute value, and so on.}
    \label{fig:stressed_power}
\end{figure}

Using the proposed method in Sections~\ref{subsec:stress_wind_speed} and \ref{subsec:stress_other_weather}, we have generated 1000 stressed weather scenarios, based on which we can calculate 1000 stressed wind power scenarios for each time period and then convert them to the probabilistic forecasting of wind power with confidence intervals (CIs). Fig.~\ref{fig:stressed_power}(a) shows the probabilistic forecasting of wind power for Jan 30, 2013 generated with the proposed weather-driven method. Due to the existence of extreme data points (e.g., when the forecast wind speed is in Region~III and the actual wind speed is in Region~IV), the possible distribution of stressed wind power covers the entire range between 0 and the rated power. 
However, when, for example, the 0.5\% extreme cases are ignored, the distribution pattern of the forecast errors becomes clearer. 
Before hour 10 of Jan 30, 2013, although the wind speeds could be relatively low, they are distributed mostly in the ``sensitive'' Region~II, where the wind power is proportional to the cube of the wind speed. 
Therefore, both positive and negative forecast errors exist, while the absolute values of the errors generally do not exceed 50\% of the rated power. 
After hour 10 of Jan 30, 2013, the forecast wind power reaches the rated power, but forecast errors still exist due to the occasional transitions of wind speed between Region~II and Region~III.

We also compare the weather-driven approach with a weather-ignorant benchmark, which relies on the historical distribution of wind power forecast errors. As in \cite{bruninx2014statistical}, the benchmark is generated as follows:

\noindent
\underline{\textit{Step 1}}: Calculate wind power forecast error $P^{E}$ by subtracting forecast wind power $P^{F}$ from actual wind power $P^{A}$.

\noindent
\underline{\textit{Step 2}}: Divide the entire wind power interval into $K$ equal-width power bins, then find the best-fit parametric distributions of $P^{E}$ when $P^{F}$ is in different power bins, i.e., $f(P^{E}|{P^{F}} \in {\rm{power \ bin}}\ k),\ \forall k \in K$.

\noindent
\underline{\textit{Step 3}}: For each time step in the test case, generate 1000 wind power forecast errors according to the conditional probability distributions obtained in Step 2, and then convert them into probabilistic wind power forecasts  with different confidence intervals, as shown in Fig.~\ref{fig:stressed_power}(b).

According to Fig.~\ref{fig:stressed_power}, the weather-ignorant benchmark results in narrower forecast error confidence intervals compared to the proposed weather-driven approach.
In this example, the actual wind power is contained in the CI=60\% confidence of the weather-driven forecast (Fig.~\ref{fig:stressed_power}(a)) while it consistently exceed the weather-ignorant probabilistic forecast even when CI=80\% (Fig.~\ref{fig:stressed_power}(b)).

\section{Flexibility Reserve Sizing and Allocation}
\label{sec:reserve}
Flexibility reserve is procured to prevent frequency excursions, load-shedding, or excessive VRES curtailment \cite{ortega2020risk}.  
We now integrate the developed weather-driven probabilistic forecasting into day-ahead planning using a security-constrained unit commitment (SCUC) model, which is typical for day-ahead operations of US ISOs. 

\subsection{Reserve Sizing Using Stressed Wind Power}
\label{susubsec:reserve_requirement}

We follow \cite{ortega2020risk} and show the application of the weather-driven uncertainty model in Section~\ref{sec:wind} and the wind power scenarios it produces for deriving extent-based, probability-based, and risk-based reserve requirements.

\subsubsection{Extent-Based Reserve Requirements}
\label{subsubsec:extent_based_reserve}
Extent-based reserve requirements are computed as:
\begin{align}
    R^{+}_t = & \varepsilon P^{F}_t \label{extent_R_up}\\
    R^{-}_t =  & \min \left\{ P^{R} - P^{F}_t, \ \varepsilon P^{F}_t \right\}, \label{extent_R_down}
\end{align}
where $R^{+}_t$ and $R^{-}_t$ are the system-wide upward and downward reserve requirements at time $t$, $P^{F}_t$ is the day-ahead point forecast of wind power at time $t$, $P^{R}$ is the rated power of the wind farm, and $\varepsilon$ is the expected extent of deviation, which is predefined by the system operator or regulatory agency. 

\subsubsection{Probability-Based Reserve Requirements}
\label{subsubsec:probability_based_reserve}
Probability-based reserve requirements are computed such that the probability of reserve excess or shortfall does not exceed a predefined limit. 
For the scenarios of stressed wind power described in Section~\ref{subsec:stress_wind_power}, which are assumed to be uniformly distributed, we can obtain the probability-based reserve requirements as follows:
\begin{align}
    R^{+}_t = &  \max \left\{0, P^{F}_t - {\tilde P}^{S}_{0.5 N (1-{\rm{CI}}),t} \right\} \label{Probability_R_up}\\
    R^{-}_t =  & \max \left\{0, {\tilde P}^{S}_{0.5 N (1+{\rm{CI}},t} -  P^{F}_t \right\}, \label{Probability_R_down}
\end{align}
where ${\tilde P}^{S}_{i,t}$ is the $i^{th}$ stressed wind power scenario at time $t$ sorted from the smallest to the largest, CI is a user-defined confidence level (i.e., probability of reserve sufficiency, $\rm{CI} \in [0,1]$), $N$ is the number of scenarios ($N=1000$ in this paper). 

\subsubsection{Risk-Based Reserve Requirements}
\label{subsubsec:risk_based_reserve}
Risk-based reserve requirements are computed such that the risk of reserve excess or shortfall does not exceed a predefined limit. 
Risk $\rho$ of a scenario is given as the expected loss of load caused by insufficient reserve defined as $\rho = P \times \xi$, where $P$ is the probability of a wind power deviation (i.e., wind power forecast error) being greater than the reserve requirement, and $\xi$ is the extent of this deviation \cite{ortega2020risk}. 
As $P$ is a conditional probability, this risk calculation is not straightforward and, as discussed in \cite{ortega2020risk}, requires integrating over the probability density function (PDF) of the wind power forecast errors. 
Using the scenarios of stressed wind power in Section~\ref{subsec:stress_wind_power}, risk calculation does not require integration but can be performed by a simpler counting procedure shown in Algorithm \ref{alg:risk_based_reserve}.
Leveraging the uniform scenario distribution, Algorithm \ref{alg:risk_based_reserve} computes risk, i.e., the expected reserve shortfall, by iteratively reducing the reserve requirement from the most extreme scenarios and counting the number of scenarios that are not covered by the reserve after the reduction. The relative number of these scenarios times the distance from the extreme cases is the risk and the algorithm finishes once the desired risk level is reached.

\begin{algorithm}[t]
    \footnotesize
    \SetAlgoLined
    \SetKwInOut{Input}{input}\SetKwInOut{Output}{output}
    \Input{stressed wind power $\{{\tilde P}^{str}_{i,t} \}_{i \in \set{S}, t \in \set{T}}$;
    forecast wind power $\{P^{F}_{t}\}_{t \in \set{T}}$; upper limit of risk $\rho$ \\
    }
    \Output {reserve requirements $\{R^{+}_{t}, R^{-}_{t}\}_{t \in \set{T}}$;
    }
    \Begin{
      \For{$t \in \set{T},\ i=1:N$}
          {
           $devi^+ = {\tilde P}^{S}_{i,t}-{\tilde P}^{S}_{1,t}$;\\
           $devi^- = {\tilde P}^{S}_{N,t}-{\tilde P}^{S}_{N-i,t}$;\\
           $risk^+ = devi^+ \times (i-1)/N $\\
           $risk^- = devi^+ \times ((N-i)-1)/N $\\
           \If{$risk^+ \le \rho$}
              {$R^{+}_{t} = P^{F}_{t} - {\tilde P}^{S}_{i,t}$}
           \If{$risk^- \le \rho$}
              {$R^{-}_{t} = {\tilde P}^{S}_{N-i,t} - P^{F}_{t}$}
            \If{$(risk^+ > \rho)$ \& $(risk^- > \rho)$}
                {\textbf{break}}
          }
      \KwRet{$\{R^{+}_{t}, R^{-}_{t}\}_{t \in \set{T}}$}
     }
    \caption{Risk-Based Reserve Sizing}
    \label{alg:risk_based_reserve}
\end{algorithm}

\begin{figure}[!t]
    \centering
    \includegraphics[width=1\linewidth]{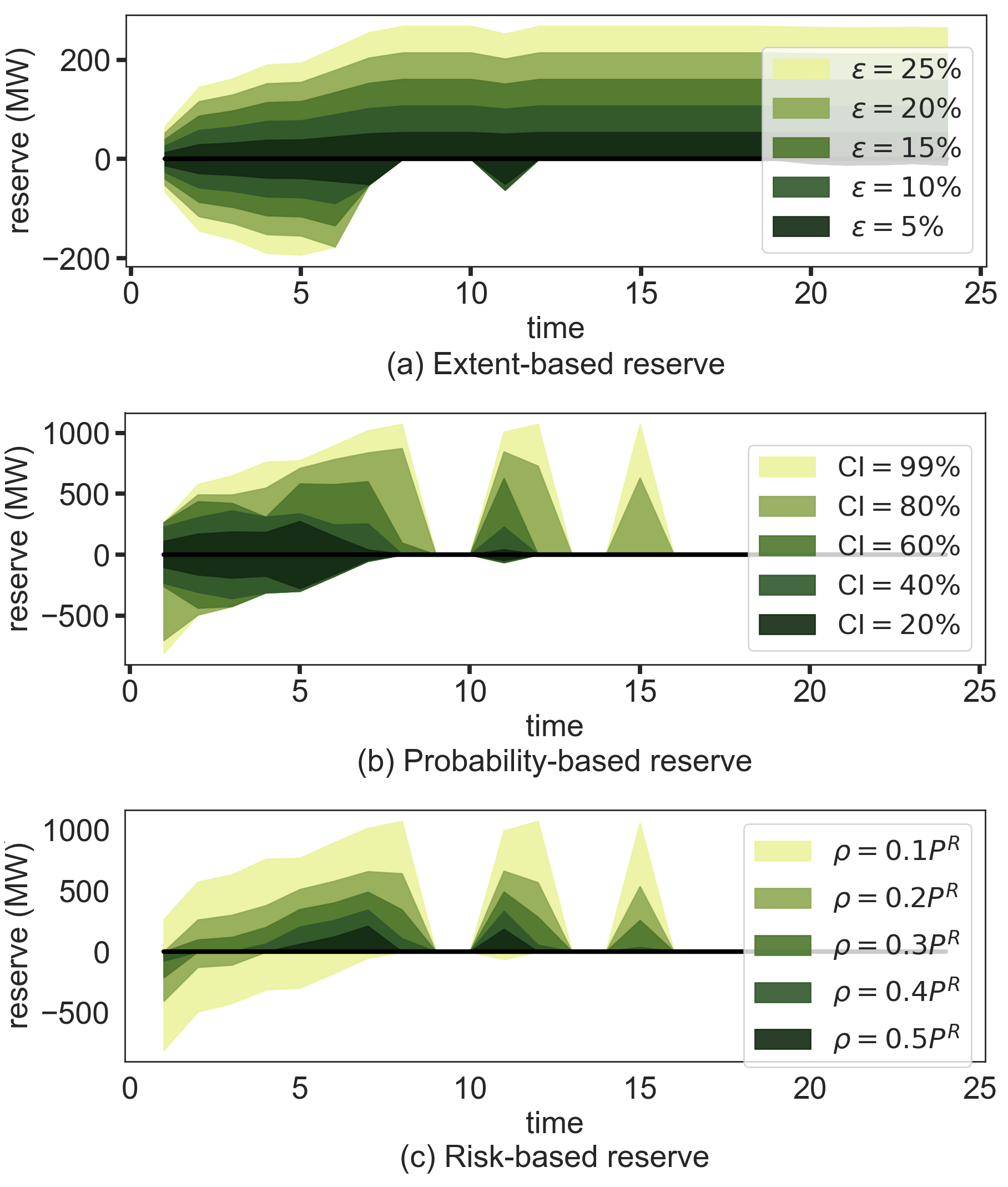}
    \caption{Resulting reserve requirements for (a) extent-based reserve with different extents $\epsilon$, (b) probability-based reserve with different confidence intervals CI, (c) and risk-based reserve with different risk-levels $\rho$, as per Eqs.~\cref{extent_R_up,extent_R_down}, Eqs.~\cref{Probability_R_up,Probability_R_down}, and Algorithm~\ref{alg:risk_based_reserve}, respectively.}
    \label{fig:three_reserve}
\end{figure}

Fig.~\ref{fig:three_reserve} shows the reserve requirements resulting from Algorithm \ref{alg:risk_based_reserve} for the three methods with different extent levels, confidence intervals, and risk levels. 
The probability-based and the risk-based reserve requirements are obtained using the stressed wind power scenarios in Fig.~\ref{fig:stressed_power}, while the extent-based reserve requirements are calculated based on the forecast wind power (red line in Fig.~\ref{fig:stressed_power}). 
It can be seen that before hour 10 of Jan 30, 2013, the trends of the three methods of quantifying reserve requirements are similar, i.e., both upward reserve and downward reserve are required.
After hour 10 of Jan 30, 2013, only upward reserve is required for most of the time. Meanwhile, the extent-based reserve requirements are always non-zero, which is overly robust and may lead to unnecessary cost increases.
The probability- and risk-based approaches relax this conservatism.

\subsection{Reserve Allocation in SCUC Model}
\label{susubsec:SCUC_model}

\subsubsection{Base SCUC Model}
\label{susubsec:base_model}
We use the SCUC formulation \cref{mod:B_scuc} from \cite{mieth2022risk} as the basis for our analysis.
Note that \cref{mod:B_scuc} considers contingency reserve, i.e., operational reserve to respond to unplanned outages of generation or transmission equipment, but it does not consider flexibility reserve in the SCUC formulation. 
\allowdisplaybreaks
\begin{subequations}
\begin{align}
\min \quad 
    & \sum\limits_{t\in\set{T}}\sum\limits_{g\in\set{G}} t_{g,t} + u_{g,t} C_{g}^{0} + v_{g,t} C_{g}^{\rm{SU}} + w_{g,t} C_{g}^{\rm{SD}} \label{B_scuc:objective}\\
\text{s.t. }\forall t\in\set{T}: \hspace{-1cm}& \nonumber \\
    & t_{g,t} \ge p_{g,t} C_{1,o,g} + C_{0,o,g}, \quad \forall g\in\set{G}, \forall{o}\in\set{O} \label{B_scuc:pwlc}\\
    & \sum\nolimits_{s=t-UT_g-1}^t v_{g,s} \leq u_{g,t} , \quad \forall g\in\set{G} \label{B_scuc:uptime}\\
    & \sum\nolimits_{s=t-DT_g-1}^t w_{g,s} \leq 1-u_{g,t} , \quad \forall g\in\set{G} \label{B_scuc:downtime}\\
    & v_{g,t} - w_{g,t} = u_{g,t} - u_{g,t-1} , \quad \forall g\in\set{G} \label{B_scuc:su_sd_indicator}\\
    & u_{g,t} P^{\min}_g \le p_{g,t}, \quad \forall g\in\set{G} \label{B_scuc:gen_lower_limit} \\
    & u_{g,t} P^{\max}_g \geq p_{g,t} + r_{g,t}^S , \quad \forall g\in\set{G} \label{B_scuc:gen_upper_limit} \\
    & p_{g,t} - p_{g,t-1} \leq R_g^{60} u_{g,t-1} + v_{g,t} P_{g}^{\min} , \ \forall g\in\set{G} \label{B_scuc:ramp_up}\\
    & p_{g,t-1} - p_{g,t} \leq R_g^{60} u_{g,t} + w_{g,t} P_{g}^{\min} , \quad \forall g\in\set{G}\label{B_scuc:ramp_down}\\
    & f_{ij,t} = B_{ij}(\theta_{i,t} - \theta_{j,t}) , \quad  \forall ij\in\set{L} \label{B_scuc:power_flows} \\
    & \theta_{ref,t} = 0  \label{B_scuc:reference} \\
    & -f^{\max}_{ij} \leq f_{ij,t} \leq f^{\max}_{ij} , \quad  \forall ij\in\set{L} \label{B_scuc:power_flow_limits}\\
    & \sum\nolimits_{g\in\set{G}_i} p_{g,t} + \sum\nolimits_{w\in\set{W}_i} p_{w,t}^{\rm{DA}} + \sum\nolimits_{j:ij\in\set{L}}f_{ij,t}  \nonumber \\
    & \hspace{0.5cm} - \sum\nolimits_{j:ji\in\set{L}}f_{ji,t} = D_{i,t}^{\rm{DA}}, \quad \forall i \in \set{N} \label{B_scuc:system_balance}\\
    & \sum\nolimits_{g\in\set{G}} r_{g,t}^S \geq p_{g,t} + r_{g,t}^S , \quad \forall g\in\set{G}\label{B_scuc:max_reserve}\\
    & \sum\nolimits_{g\in\set{G}} r_{g,t}^S \geq R^D \sum\nolimits_{i\in\set{N}} D_{i,t}^{\rm{DA}} \label{B_scuc:min_reserve}\\
    & r_{g,t}^{S} \le R_{g}^{10} , \quad \forall g\in\set{G} \label{B_scuc:spinning_ramp}\\
    & u_{g,t} \in \{0,1\}, \quad \forall g\in\set{G} \label{B_scuc:u_definiton} \\
    &  0 \le v_{g,t}, w_{g,t} \leq 1, \quad \forall g\in\set{G}. \label{B_scuc:v_w_definiton} 
\end{align}%
\label{mod:B_scuc}%
\end{subequations}%
\allowdisplaybreaks[0]%
Objective \cref{B_scuc:objective} minimizes the system cost using no-load costs $C_g^0$, start-up costs $C_g^{\rm{SU}}$, shut-down costs $C_g^{\rm{SD}}$, and piece-wise linear generator cost functions defined in \cref{B_scuc:pwlc}, where $\mathcal{G}$ is the set of conventional generators, $\mathcal {O}$ is the set of linear cost curve segments for conventional generators, $p_{g,t}$ is the power output of generator $g$ at time $t$, $C_{0,o,g}$ and $C_{1,o,g}$ are the constant and linear cost coefficients of the operating cost for generator $g$ in cost segment $o$, respectively.
Constraints \cref{B_scuc:uptime,B_scuc:downtime,B_scuc:su_sd_indicator} relate binary variables $u_{g,t}$, $v_{g,t}$ and $w_{g,t}$ that denote commitment, start-up and shut-down decisions.  
Commitment changes are restricted by the minimum up- and down-time limits enforced in \cref{B_scuc:uptime,B_scuc:downtime}, where $DT_g$ and $UT_g$ are the minimum downtime (off) and the minimum uptime (on) of generator $g$. 
Note that it is sufficient to explicitly define $u_{g,t}$ as binary in \cref{B_scuc:u_definiton}, while $v_{g,t}$ and $ w_{g,t}$ are continuous within interval $[0,1]$ as in \cref{B_scuc:v_w_definiton}. 
Capacity limits of generators are enforced in \cref{B_scuc:gen_lower_limit,B_scuc:gen_upper_limit}, where $r^S_{g,t}$ is the spinning reserve provided by generator $g$ at time $t$, and $P_{g}^{\max}$ and $P_{g}^{\min}$ are the maximum and minimum power output for generator $g$, respectively.
Constraints \cref{B_scuc:ramp_up,B_scuc:ramp_down} enforce generator ramping limits, where $R^{60}_{g}$ is the 60-min ramp rate for generator $g$.
The DC power flow equations, reference bus definition and thermal power flow limits are modeled as in \cref{B_scuc:power_flows,B_scuc:reference,B_scuc:power_flow_limits}, where $\mathcal L$ is the set of lines, $B_{ij}$ is the susceptance of line $ij$, $\theta_{i,t}$ is the voltage angle at node $i$ at time $t$.
Eq.~\cref{B_scuc:system_balance} ensures the nodal power balance by accounting for the generation, demand, and power flows at all nodes in set $\mathcal N$. Here, $\set{W}$ is the set of wind farms, $p_{w,t}^{\rm{DA}}$ is the day-ahead forecast wind power of wind farm $w$, $D_{i,t}^{\rm{DA}}$ is the day-ahead forecast load at node $i$. 
Finally, \cref{B_scuc:min_reserve,B_scuc:max_reserve,B_scuc:spinning_ramp} enforce contingency reserve requirements. Specifically, the total contingency reserve must cover at least the power loss in case of the largest generator outage, \cref{B_scuc:max_reserve}, or a fraction $R^D$ of system demand, \cref{B_scuc:min_reserve}, and are limited by the 10-min ramp rate of $R_{g}^{10}$.
    
\subsubsection{System-Wide Flexibility Reserve}
\label{susubsec:system_level_reserve}
We first add system-wide flexibility reserve requirements to \cref{mod:B_scuc}, which ignores network limits and therefore may not guarantee flexibility deployment in real time, as follows: 
\allowdisplaybreaks
\begin{subequations}
\begin{align}
\min \quad 
    & \sum\limits_{t\in\set{T}}\sum\limits_{g\in\set{G}} t_{g,t} + u_{g,t} C_{g}^{0} + v_{g,t} C_{g}^{\rm{SU}} + w_{g,t} C_{g}^{\rm{SD}} \label{S_scuc:objective}\\
\text{s.t. } & \forall t\in\set{T}: \cref{B_scuc:pwlc}-\cref{B_scuc:su_sd_indicator}, \cref{B_scuc:power_flows}-\cref{B_scuc:v_w_definiton}\nonumber \\
    & r_{g,t}^{+} \le R_{g}^{10} , \quad \forall g\in\set{G} \label{S_scuc:flexibility_upward_ramp}\\
    & r_{g,t}^{-} \le R_{g}^{10} , \quad \forall g\in\set{G} \label{S_scuc:flexibility_downward_ramp}\\
    & u_{g,t} P^{\min}_g \le p_{g,t}  - r_{g,t}^{-}, \quad \forall g\in\set{G} \label{s_scuc:gen_lower_limit} \\
    & u_{g,t} P^{\max}_g \geq p_{g,t}  + r_{g,t}^{+} + r_{g,t}^{S}, \quad \forall g\in\set{G} \label{s_scuc:gen_upper_limit} \\
    & p_{g,t} + r_{g,t}^{+} - p_{g,t-1} + r_{g,t-1}^{-} \leq R_g^{60} u_{g,t-1} + v_{g,t} P_{g}^{\min}, \nonumber \\
    & \hspace{5cm} \forall g\in\set{G} \label{s_scuc:ramp_up}\\
    & p_{g,t-1}  + r_{g,t-1}^{+} - p_{g,t} + r_{g,t}^{-} \leq R_g^{60} u_{g,t} + w_{g,t} P_{g}^{\min},\nonumber \\
    & \hspace{5cm} \forall g\in\set{G}\label{s_scuc:ramp_down}\\
    & \sum\nolimits_{g\in\set{G}} r_{g,t}^{+} \geq R_t^{+} +  \sum\nolimits_{i\in\set{N}} \epsilon^d D_{i,t}^{\rm{DA}}, \label{s_scuc:total_upward_reserve} \\
    & \sum\nolimits_{g\in\set{G}} r_{g,t}^{-} \geq R_t^{-} +  \sum\nolimits_{i\in\set{N}} \epsilon^d D_{i,t}^{\rm{DA}}, \label{s_scuc:total_downward_reserve}
\end{align}%
\label{mod:s_scuc}%
\end{subequations}%
\allowdisplaybreaks[0]%
where $r_{g,t}^{+}$ and $r_{g,t}^{-}$ are the upward and downward flexibility reserve provided by generator $g$ at time $t$. As per \cref{S_scuc:flexibility_upward_ramp,S_scuc:flexibility_downward_ramp}, flexibility reserve must be available within 10 minutes. The total flexibility reserve requirement consists of two parts, namely the reserve for wind ($R_t^{+}$ and $R_t^{-}$) and the reserve for load (extent-based, where $\epsilon^d$ is expected extent of deviation).
Note that we model flexibility and contingency reserve separately such that adding flexibility reserve will not affect contingency reserve requirements.

\subsubsection{Zonal Flexibility Reserve}
\label{susubsec:zonal_reserve}
To increase flexibility deliverability and avoid network congestion, most power systems allocate reserve requirements among zones based on their specific needs and congestion patterns.
For example, the NYISO system has 11 zones 
and reserve requirements for different zones are defined separately \cite{NYISO_zonal_reserve}. 
Therefore, instead of system-wide reserve requirements, in the following model we consider zonal reserve requirements, where $\set{N^A}$ is the set of zones, $\set{D}_a$ is the sets of loads in zone $a$, $R_{a,t}^{+}$ and $R_{a,t}^{-}$ are the upward and downward flexibility reserve requirements for zone $a$, respectively.
\allowdisplaybreaks
\begin{subequations}
\begin{align}
\min \quad 
    & \sum\limits_{t\in\set{T}}\sum\limits_{g\in\set{G}} t_{g,t} + u_{g,t} C_{g}^{0} + v_{g,t} C_{g}^{\rm{SU}} + w_{g,t} C_{g}^{\rm{SD}} \label{z_scuc:objective}\\
\text{s.t. } & \forall t\in\set{T}: \cref{B_scuc:pwlc}-\cref{B_scuc:su_sd_indicator}, \cref{B_scuc:power_flows}-\cref{B_scuc:v_w_definiton}, \cref{S_scuc:flexibility_upward_ramp}-\cref{s_scuc:ramp_down}\nonumber \\
    & \sum\nolimits_{g\in\set{G}_a}\!r_{g,t}^{+}\! \ge \!R_{a,t}^{+} \!+\!  \sum\nolimits_{i\in\set{D}_a} \!\! \epsilon^d D_{i,t}^{\rm{DA}},\ \forall a \in \set{N^A}  \label{z_scuc:upward_reserve_balance}\\
    & \sum\nolimits_{g\in\set{G}_a}\! r_{g,t}^{-} \!\ge \! R_{a,t}^{-}\! +\!  \sum\nolimits_{i\in\set{D}_a}\!\! \epsilon^d D_{i,t}^{\rm{DA}},\ \forall a \in \set{N^A}\!\!. \label{z_scuc:downward_reserve_balance}
\end{align}%
\label{mod:z_scuc}%
\end{subequations}%
\allowdisplaybreaks[0]%

\subsubsection{Nodal Flexibility Reserve}
\label{susubsec:nodal_reserve}
While zonal reserve requirements can avoid some congestion effects, they are typically defined in a static and long-term manner and ignore changing system conditions and intra-zonal congestion. 
To address this shortcoming, we formulate a nodal reserve requirement as:
\allowdisplaybreaks
\begin{subequations}
\begin{align}
\min \quad 
    & \sum\limits_{t\in\set{T}}\sum\limits_{g\in\set{G}} t_{g,t} + u_{g,t} C_{g}^{0} + v_{g,t} C_{g}^{\rm{SU}} + w_{g,t} C_{g}^{\rm{SD}} \label{n_scuc:objective}\\
\text{s.t. } & \forall t\in\set{T}: \cref{B_scuc:pwlc}-\cref{B_scuc:su_sd_indicator}, \cref{B_scuc:power_flows}-\cref{B_scuc:v_w_definiton}, \cref{S_scuc:flexibility_upward_ramp}-\cref{s_scuc:ramp_down}\nonumber \\
    & f_{ij,t}^{+} = B_{ij}( \theta_{i,t}^{+} - \theta_{j,t}^{+}), \quad  \forall ij\in\set{L} \label{n_scuc:upward_reserve_flows} \\
    & \sum\nolimits_{g\in\set{G}_i} r_{g,t}^{+} + \sum\nolimits_{j:ij\in\set{L}} f_{ij,t}^{+} - \sum\nolimits_{j:ji\in\set{L}} f_{ji,t}^{+}  \nonumber \\
    & \hspace{0.5cm} = R_{i,t}^{+}  + \epsilon^d D_{i,t}^{\rm{DA}},  \quad \forall i \in \set{N} \label{n_scuc:upward_reserve_balance}\\
    & f_{ij,t}^{-} = B_{ij}( \theta_{i,t}^{-} - \theta_{j,t}^{-}), \quad  \forall ij\in\set{L} \label{n_scuc:downward_reserve_flows} \\
    & \sum\nolimits_{g\in\set{G}_i} r_{g,t}^{-} + \sum\nolimits_{j:ij\in\set{L}} f_{ij,t}^{-} - \sum\nolimits_{j:ji\in\set{L}} f_{ji,t}^{-}  \nonumber \\
    & \hspace{0.5cm}  = R_{i,t}^{-} + \epsilon^d D_{i,t}^{\rm{DA}},\quad \forall i \in \set{N} \label{n_scuc:downward_reserve_balance}\\
    & \theta_{ref,t}^{+} = 0,\ \theta_{ref,t}^{-} = 0 \label{n_scuc:reference_bus} \\
    & -f^{\max}_{ij} \leq f_{ij,t} \leq f^{\max}_{ij}, \ \ \forall ij\in\set{L} \label{n_scuc:power_flow_limits} \\
    & -f^{\max}_{ij} \leq f_{ij,t} + f_{ij,t}^{+} \leq f^{\max}_{ij}, \ \ \forall ij\in\set{L} \label{n_scuc:power_flow_limits_up} \\
    & -f^{\max}_{ij} \leq f_{ij,t} + f_{ij,t}^{-} \leq f^{\max}_{ij}, \ \ \forall ij\in\set{L},\label{n_scuc:power_flow_limits_down}
\end{align}%
\label{mod:n_scuc}%
\end{subequations}%
\allowdisplaybreaks[0]%
where we assume that every wind farm is connected to one node in the system. Variables $f_{ij,t}^{+}$, $f_{ij,t}^{-}$ and $\theta_{i,t}^{+}$, $\theta_{i,t}^{-}$ are the additional flows and voltage angle changes due to reserve deployment.
Since we use a linear DC power flow model, the power flow and additional power flow caused by reserve deployment can be superimposed such that $f_{ij,t} + f_{ij,t}^{+}$ is the cumulative flow between node $i$ and $j$ at time $t$ when the upward reserve is deployed.  

\subsection{Reserve Deployment in Real-time Dispatch Model}
\label{susubsec:RT_model}
If the real-time (RT) wind power is different from the day-ahead (DA) forecast wind power, then generators need to provide upward or downward flexibility during RT scheduling. 
The dispatched RT flexibility of generator $g$ at time $t$ can be calculated as $R_{g,t} = P^{\rm{RT}}_{g,t} - P^{\rm{DA}}_{g,t}$, where $P^{\rm{RT}}_{g,t}$ and $P^{\rm{DA}}_{g,t}$ are the output power of generator $g$ at time $t$ during RT and DA scheduling, respectively. 
We evaluate the actual reserve dispatch, $R_{g,t}$, in terms of the scheduled reserve, $r_{g,t}^{+}$ and $r_{g,t}^{-}$, in the DA stage:

\noindent
\textbf{Range I}: If $0 \le R_{g,t} \le r_{g,t}^{+}$ or $ 0 \ge R_{g,t} \ge -r_{g,t}^{-}$, the dispatched flexibility has been scheduled at the DA stage. 

\noindent
\textbf{Range II}: If $R_{g,t} > r_{g,t}^{+}$ or $ -R_{g,t} \le r_{g,t}^{-}$, then the excess flexibility, i.e., $R_{g,t}-r_{g,t}^{+}$ (if $R_{g,t}>0$) or $-R_{g,t}- r_{g,t}^{-}$ (if $R_{g,t}<0$), is an impromptu emergency response and has not been scheduled at the DA stage. 

The cost for flexibility in Range I are assumed to be smaller or equal to the cost for flexibility in Range II to account for higher cost of very-short term changes in production levels beyond a planned interval.  
If the system cannot be balanced from available reserve, load-shedding or wind-spillage occurs.
We model the RT scheduling procedure as:
\allowdisplaybreaks
\begin{subequations}
\begin{align}
\min \quad 
    & \sum\nolimits_{t\in\set{T}} M_t^{\rm{PG}} + M_t^{\rm{LS}} + M_t^{\rm{WS}} + M_t^{\rm{RD}} \label{RT:objective}\\
\text{s.t. } & \forall t\in\set{T}: \cref{B_scuc:pwlc}-\cref{B_scuc:su_sd_indicator}, \cref{B_scuc:ramp_up}-\cref{B_scuc:power_flow_limits}, \cref{B_scuc:u_definiton} \nonumber \\
    & M_t^{\rm{PG}} = \sum\nolimits_{g\in\set{G}} t_{g,t} + u_{g,t} C_{g}^{0} + v_{g,t} C_{g}^{\rm{SU}} + w_{g,t} C_{g}^{\rm{SD}} \label{RT:energy_cost}\\
    & M_t^{\rm{LS}} =  \sum\nolimits_{i\in\set{N}} \tilde D_{i,t} C_i^{\rm{LS}} \label{RT:load_shedding}\\
    & M_t^{\rm{WS}} = \sum\nolimits_{w\in\set{W}} \tilde p_{w,t} C_w^{\rm{WS}} \label{RT:wind_spillage}\\
    & M_t^{\rm{RD}} = \sum\nolimits_{g\in\set{G}}(C^{\text{RD-I}}_{g} R_{g,t}^{\text{I}} + C^{\text{RD-II}}_{g} R_{g,t}^{\text{II}}) \label{RT:redispatch_cost}\\
    & u_{g,t} P^{\min}_g \le p_{g,t} \le u_{g,t} P^{\max}_g, \quad \forall g\in\set{G} \label{RT:gen_limit} \\
    & \sum\nolimits_{g\in\set{G}_i} p_{g,t} {\rm{+}} \sum\nolimits_{w\in\set{W}_i} (p_{w,t}^{\rm{RT}}- \tilde p_{w,t}) {\rm{+}} \sum\nolimits_{j:ij\in\set{L}}f_{ij,t}  \nonumber \\
    & \hspace{0.5cm} - \sum\nolimits_{j:ji\in\set{L}}f_{ji,t} = D_{i,t}^{\rm{RT}}  - \tilde D_{i,t} , \quad \forall i \in \set{N} \label{RT:system_balance}\\
    & R_{g,t}^{\text{I}} = \max \Big\{0,\ \min\{P_{g,t} - P_{g,t}^{\rm{DA}},\ r_{g,t}^{+}\}, \nonumber \\
    & \hspace{0.5 cm} \min\{P_{g,t}^{\rm{DA}} - P_{g,t},\ r_{g,t}^{-}\} \Big\}, \quad \forall{g}\in\set{G} \label{RT:low_balancing}\\
    & R_{g,t}^{\text{II}} = \max \big\{0, \ P_{g,t} {\rm{-}} P_{g,t}^{\rm{DA}} {\rm{-}} r_{g,t}^{+},\ P_{g,t}^{\rm{DA}} {\rm{-}} P_{g,t} {\rm{-}} r_{g,t}^{-} \big\} \nonumber \\  
    & \hspace{0.5cm} \forall{g}\in\set{G}
    \label{RT:high_balancing}\\
    & u_{g,t} = u_{g,t}^{\rm{DA}}, \ v_{g,t} = v_{g,t}^{\rm{DA}}, \ w_{g,t} = w_{g,t}^{\rm{DA}} , \quad \forall g\in\set{G}, \label{RT:u_v_w}
\end{align}%
\label{mod:RT}%
\end{subequations}%
\allowdisplaybreaks[0]%
where $M_t^{\rm{PG}}$, $M_t^{\rm{LS}}$, $M_t^{\rm{WS}}$, $M_t^{\rm{RD}}$ are the cost of power generation, load shedding, wind spillage and redispatch at time $t$, $D_{i,t}^{\rm{RT}}$ is the real-time load, $C_i^{\rm{LS}}$ and $C_w^{\rm{WS}}$ are the penalties for load shedding and wind spillage, $C_{g}^{\text{RD-I}}$ and $C_{g}^{\text{RD-II}}$ are the unit cost of generator $g$ providing flexibility in Ranges I and II, respectively, $p_{w,t}^{\rm{RT}}$ is the real-time wind power, $\tilde D_{i,t}$ is the curtailed load at node $i$, and $\tilde p_{w,t}$ is the curtailed wind power of wind farm $w$. 
In \cref{mod:RT}, the commitment status ($u_{g,t}$, $v_{g,t}$, $w_{g,t}$) is a fixed parameter given by DA SCUC.
   
\section{Case Study}
\label{sec:Cases}

\subsection{Data Resource and Simulation Environment}
\begin{figure}[!t]
    \centering
    \includegraphics[width=1\linewidth]{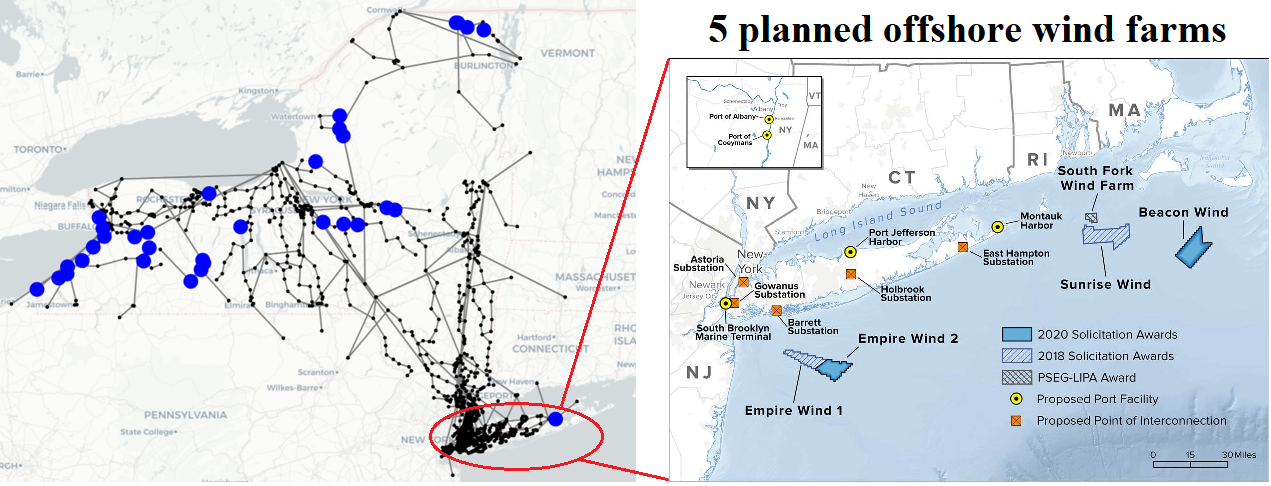}
    \caption{Synthetic NYISO system. On the left, the black lines are transmission lines, the small black dots are buses, and the large blue dots are the 33 existing onshore wind farms. On the right, the blocks shaded or filled with blue are the 5 planned offshore wind farms near Long Island, NY.}
    \label{fig:wind_farms}
\end{figure}

We conduct numerical experiments using a NYISO system model with 11 zones, 1819 buses, 2207 lines, 362 generators and 38 wind farms (including 33 existing onshore wind farms and 5 planned offshore wind farms that are still under construction), as shown in Fig.~\ref{fig:wind_farms}.
For this case study we use data from February 2013.
As in Section~\ref{subsec:weather_data}, we use weather data from the NREL WIND Toolkit \cite{Wind_toolkit}.
Additionally, we use load data from the NYISO data platform \cite{nyiso_data}. 
We generate the reserve requirements for each wind farm independently and then combine them into zonal (or system-wide) reserve requirements by adding the reserve requirements in each zone (or the whole system). 
The penalties for load shedding ($C_i^{\rm{LS}}$ in \cref{RT:load_shedding}) and wind spillage ($C_w^{\rm{WS}}$ in \cref{RT:wind_spillage}) are set to \unit[10,000]{\$/MW} and \unit[100]{\$/MW} \cite{nyisovoll}, respectively, and the unit cost for providing flexibility ($C_{g}^{\text{RD-I}}$ and $C_{g}^{\text{RD-II}}$ in \cref{RT:redispatch_cost}) are set to \unit[2]{\$/MW} and \unit[5]{\$/MW}, which captures average regulation market prices at NYISO \cite{patton2021} and prioritizes the utilization of reserve that have been scheduled in the DA phase as discussed in Section~\ref{susubsec:RT_model}.
If reserve requirements cannot be met, they are relaxed with a penalty of \unit[500]{\$/MW} corresponding to the NYISO regulation demand curve \cite{nyisovoll}.

All simulations were implemented in Python v3.8 using the Gurobi solver. All experiments were performed on a standard PC workstation with an Intel i9 processor and 16 GB RAM. 
The average computing time for the SCUC models with system-wide, zonal, and nodal reserve requirements were about 1 minute, 4 minutes and 6 minutes, while each RT model was solved in less than 6 minutes.

\subsection{Performance Evaluation}

\subsubsection{Comparing Weather-Driven Reserve with Weather-Ignorant Benchmark}
To compare the proposed weather-driven method with the weather-ignorant benchmark, we generate two sets of system-wide risk-based reserve requirements ($\rho<0.3 P^{\rm rate}$) based on the weather-driven and the weather-ignorant probabilistic forecasts and compare system cost performance in these two cases as shown in Table~\ref{tab:cost_benchmark}. The weather-driven case outperforms the weather-ignorant case, especially in terms of the RT load shedding cost and RT wind spillage cost. Consistent with the observation in Fig.~\ref{fig:stressed_power}, this is because the weather-ignorant case underestimates the required flexibility.

\begin{table}[!t]
    \caption{Costs Under Different System-wide Risk-based Reserve \tnote{1}}
    \begin{center}
    \begin{threeparttable}
        \begin{tabular}{ccc}
            \toprule
                Cost & Weather-driven & Weather-ignorant \\ 
            \midrule
                SCUC generation cost & 298.7242  & 298.4155  \\ 
                RT generation cost & 3885.4178  & 3881.6968  \\ 
                RT load shedding cost & 28.1380  & 31.3236  \\ 
                RT wind spillage cost & 65.1999  & 68.4047  \\ 
                RT redispatch cost & 5.0818  &  6.4886 \\ 
                RT total cost & 3983.8376  & 3987.9137 \\ 
            \bottomrule
        \end{tabular}%
        \begin{tablenotes}
            \item[1] In Million \$.
        \end{tablenotes}
    \end{threeparttable}
    \end{center}
    \label{tab:cost_benchmark}
\end{table}

\subsubsection{Comparing Different Levels of Reserve Requirements}
\begin{figure}[!t]
    \centering
    \includegraphics[width=1\linewidth]{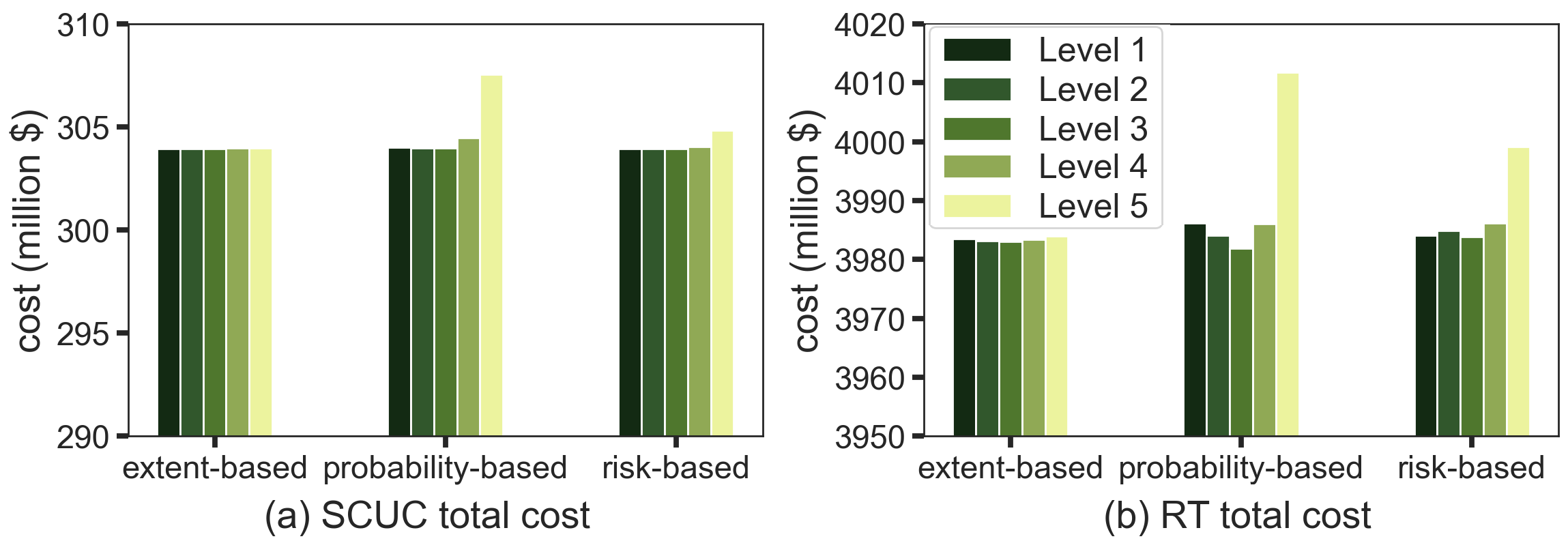}
    \caption{SCUC and RT cost at different levels of system-wide reserve. The levels are defined in Table~\ref{tab:five_levels}.}
    \label{fig:5level}
\end{figure}

We first compare five levels of reserve requirements as in Table~\ref{tab:five_levels} based on the SCUC model in \cref{mod:s_scuc} with the system-wide reserve requirements to be the benchmark in this case study. 
Fig.~\ref{fig:5level} shows the resulting costs. 
Considering SCUC costs, reserve of level 1-3 render similarly good results; considering RT costs, level 3 is the best choice for the three types of reserve (extent-, probability-, and risk-based).

\begin{table}[!t]
  \centering
  \caption{Setting of Five Levels in Different Cases}
    \begin{tabular}{cccc}
    \toprule
        Levels & Extent-based & Probability-based & Risk-based \\ 
        \midrule
        1 & $\epsilon=5\%$ & CI$=20\%$ & $\rho<0.1 P^{\rm rate}$ \\
        2 & $\epsilon=10\%$ & CI$=40\%$ & $\rho<0.2 P^{\rm rate}$ \\
        3 & $\epsilon=15\%$ & CI$=60\%$ & $\rho<0.3 P^{\rm rate}$ \\
        4 & $\epsilon=20\%$ & CI$=80\%$ & $\rho<0.4 P^{\rm rate}$ \\
        5 & $\epsilon=25\%$ & CI$=99.9\%$ & $\rho<0.5 P^{\rm rate}$ \\
    \bottomrule
    \end{tabular}%
  \label{tab:five_levels}
\end{table}

\subsubsection{Comparing System-Wide, Zonal and Nodal Reserve}

\begin{figure}[!t]
    \centering
    \includegraphics[width=1\linewidth]{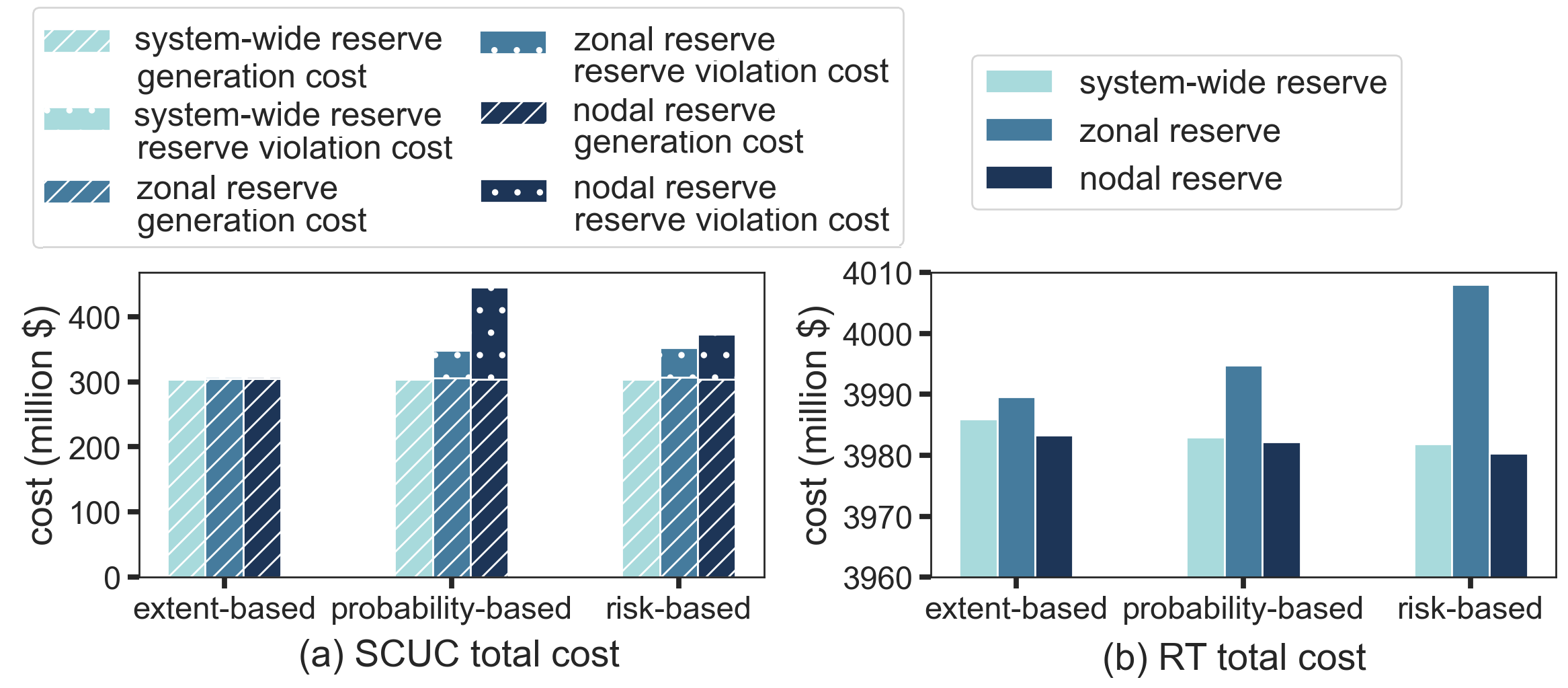}
    \caption{SCUC and RT cost of three reserve procurement models and three types of reserve.}
    \label{fig:3model_3reserve}
\end{figure}

Fig.~\ref{fig:3model_3reserve} compares the results of the system-wide, zonal and nodal reserve procurement. 
The nodal reserve results in the least RT total cost because they provide spatially more accurate reserve allocation in the system. However, the nodal reserve also corresponds to a high reserve violation penalty.
This implies that the nodal reserve sizing and allocation is only partially procured due to binding line congestion and/or insufficient local generation capacity. 
Meanwhile, compared with the system-wide reserve, the zonal reserve cannot reduce the RT cost despite its more refined allocation information. This is because sometimes the conventional generators in a given zone are not capable of scheduling enough flexibility reserve for the wind farms in that zone. 

\subsubsection{Comparing Extent-, Probability-, and Risk-based Reserve}

Since the nodal reserve corresponds to the least RT cost, we use nodal reserve as the benchmark to compare the results of the extent-based, probability-based, and risk-based reserve requirements, as shown in Table~\ref{tab:cost_different_reserve}. 
The performance of these reserve requirements is similar considering the SCUC and RT power generation cost and RT wind spillage cost, while the risk-based reserve performs better in terms of RT load shedding cost and redispatch cost. 

\begin{figure}[!t]
    \centering
    \includegraphics[width=1\linewidth]{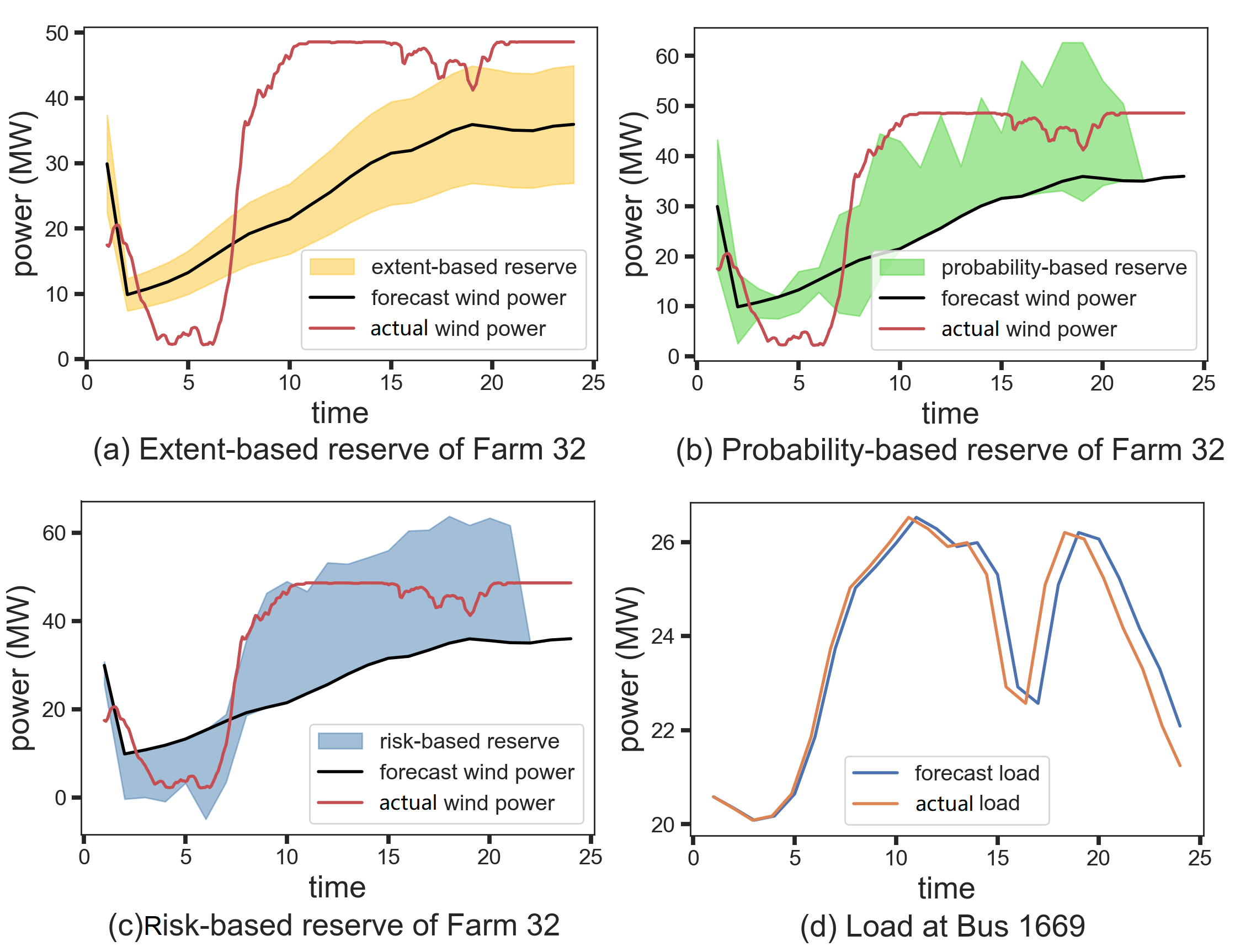}
    \caption{Three types of nodal reserve requirements for wind farm 32 and the corresponding nodal load on Feb. 27.}
    \label{fig:reserve_Feb_22}
\end{figure}

We choose the day when the difference in the RT cost of three types of reserve is the greatest (i.e., Feb. 27) as an example to show how risk-based reserve can reduce load shedding costs.
Load shedding on this day mainly occurs at and around a single bus (\#1669) which hosts a wind farm (\#32) with an installed capacity of \unit[48.6]{MW}, indicating a lack of upward reserve when the actual wind power falls short of its forecast.
Fig.~\ref{fig:reserve_Feb_22} shows the forecast and actual wind power of this wind farm on Feb. 27, together with its resulting reserve requirements.
We observe in Fig.~\ref{fig:reserve_Feb_22}(a) that the extent-base reserve requirement underestimates the required flexibility. The probability- and risk-based approaches in Fig.~\ref{fig:reserve_Feb_22}(b) and (c) improve on that and meet the requirements for the real-time injections more closely.
Note that electricity demand forecast errors, as shown in Fig.~\ref{fig:reserve_Feb_22}(d), are negligible for reserve activation and load-shedding compared to wind power forecast errors.

To analyze the redispatch cost, we define upward and downward reserve activation factors ${RAF}^{+}$ and ${RAF}^{-}$ as: 
\begin{align}
    {RAF}^{+} = & (P_{g,t}^{\rm{RT}}-P_{g,t}^{\rm{DA}})/r_{g,t}^{+} \label{RAF_up}\\
    {RAF}^{-} = & (P_{g,t}^{\rm{DA}}-P_{g,t}^{\rm{RT}})/r_{g,t}^{-}. \label{RAF_down}
\end{align}
Recall that $P_{g,t}^{\rm{DA}}$ and  $P_{g,t}^{\rm{RT}}$ are the outputs of generator $g$ at time $t$ during DA and RT scheduling, and $r_{g,t}^{+}$ and  $r_{g,t}^{-}$ are the scheduled upward and downward reserve. These reserve activation factors reflect how much DA scheduled reserve are deployed during RT scheduling, so the higher the reserve activation factors are, the more efficient are the reserve requirements. 
Table~\ref{tab:RAF} shows the resulting reserve activation factors. 
Both ${RAF}^{+}$ and ${RAF}^{-}$ of the risk-based reserve are the highest compared to the other two reserve types, which explains the lower corresponding redispatch cost.

\begin{table}[!t]
    \caption{Costs Under Different Nodal Reserve Requirements At Level 3 \tnote{1}}
    \begin{center}
    \begin{threeparttable}
        \begin{tabular}{cccc}
        \toprule
            Cost & Extent-based & Probability-based & Risk-based \\ 
        \midrule
            SCUC generation & 303.9554  & 303.9621  & 303.9670  \\ 
            RT generation & 3885.6175  & 3885.5296  & 3885.3490  \\ 
            RT load shedding & 28.1184  & 27.6284  & 26.3516  \\ 
            RT wind spillage & 65.2022  & 65.1760  & 65.1902  \\ 
            RT redispatch & 4.0662  &  3.9678 &  3.5985\\ 
            RT total & 3983.0042  & 3982.3017  & 3980.4894  \\ 
        \bottomrule
        \end{tabular}%
        \begin{tablenotes}
            \item[1] In Million \$.
        \end{tablenotes}
    \end{threeparttable}
    \end{center}
    \label{tab:cost_different_reserve}
\end{table}

\begin{table}[]
    \centering
    \caption{RAF Under Different Nodal Reserve Requirements}
    \begin{tabular}{cccc}
    \toprule
        RAF & Extent-based & Probability-based & Risk-based \\
    \midrule
        ${RAF}^{+}$ & 2.1431  & 2.3531  & 2.3800  \\ 
        ${RAF}^{-}$ & 6.7402  &  8.2036 & 8.8925  \\ 
        Total & 8.8832  & 10.5567  & 11.2725 \\ 
    \bottomrule
    \end{tabular}%
  \label{tab:RAF}
\end{table}

\section{Conclusion}
\label{sec:conclusion}
In this paper we proposed an effective method to sample wind power forecast errors from weather forecasting and its historical statistics.
We have leveraged properties of the wind power curve to derive conditional probability distributions for wind speed forecast errors. 
Using principal component analysis (PCA) we then created statistically consistent weather scenarios that include not only wind speed but also other relevant features.
Second, we demonstrated how the stressed wind power scenarios can be used for sizing and allocation flexibility reserve. 
We compared different reserve polices and showed an efficient approach for a risk-based quantification of reserve requirements.
We have applied these methods in a case study using real-world historical weather data and a realistic model of the New York ISO power system, and we demonstrate the real-time cost saving advantage of the proposed weather-driven reserve relative to the  weather-ignorant benchmark.

\bibliographystyle{IEEEtran}
\bibliography{literature}

\end{document}

%% file: math_and_tools.tex
\usepackage{amsmath,amsfonts,amsthm,amssymb}
\usepackage[bb=boondox]{mathalfa}
\usepackage[algoruled]{algorithm2e}
\usepackage{array}

\usepackage[noabbrev,capitalize]{cleveref}

\crefname{equation}{}{}
\Crefname{equation}{}{}

\theoremstyle{definition} 

\theoremstyle{plain} 

\theoremstyle{remark} 

\usepackage{tikz}

\newcommand{\set}[1]{\mathcal{#1}} 

\newcommand{\RNum}[1]{\uppercase\expandafter{\romannumeral #1\relax}}
\newcommand{\rNum}[1]{\expandafter\romannumeral #1}

\makeatletter
\newcommand{\pushright}[1]{\ifmeasuring@#1\else\omit\hfill$\displaystyle#1$\fi\ignorespaces}
\newcommand{\pushleft}[1]{\ifmeasuring@#1\else\omit$\displaystyle#1$\hfill\fi\ignorespaces}
\makeatother

\newcolumntype{C}[1]{>{\centering\arraybackslash}p{#1}} 

%% file: set_margins.tex
\usepackage{placeins}
\newcommand{\subparagraph}{}
\usepackage{titlesec}

\titlespacing*{\section}{0pt}{5pt}{3pt}
\titlespacing*{\subsection}{0pt}{3pt}{0pt}
\titlespacing*{\subsubsection}{0pt}{3pt}{0pt}